\documentclass[11pt]{article}

\usepackage{amsmath}
\usepackage{graphicx}
\usepackage{indentfirst}
\usepackage{amssymb}
\usepackage{cite}
\usepackage{color}
\usepackage{subfigure}
\usepackage{varwidth}
\usepackage{subfigure}
\usepackage[colorlinks=true, linkcolor=red, citecolor=blue, urlcolor=magenta]{hyperref}
\usepackage{xcolor}

\begin{document}

\title{Observational signatures and polarized images of rotating charged black holes in Kalb-Ramond Gravity}

\date{}
\maketitle

\begin{center}
\author{Chen-Yu Yang,}$^{a}$\footnote{E-mail: chenyuyang\_2024@163.com}
\author{Ke-Jian He,}$^{a}$\footnote{E-mail: kjhe94@163.com}
\author{Xiao-Xiong Zeng,}$^{b}$\footnote{E-mail: xxzengphysics@163.com}
\author{Li-Fang Li}$^{c}$\footnote{E-mail: lilifang@imech.ac.cn (Corresponding author)}
\\

\vskip 0.25in
$^{a}$\it{Department of Mechanics, Chongqing Jiaotong University, Chongqing 400000, People's Republic of China}\\
$^{b}$\it{College of Physics and Electronic Engineering, Chongqing Normal University, Chongqing 401331, People's Republic of China}\\
$^{c}$\it{Center for Gravitational Wave Experiment, National Microgravity Laboratory, Institute of Mechanics, Chinese Academy of Sciences, Beijing 100190, People's Republic of China}

\end{center}

\vskip 0.6in
{\abstract
{This paper investigates the shadow images of rotating charged black holes in Kalb–Ramond (KR) gravity, using an accretion disk that is both optically and geometrically thin and located on the equatorial plane as the light source model. The results show that, compared to Sgr A*, the observational data of M87* impose stronger constraints on the charge parameter $Q$ and the Lorentz-violating parameter $\mathcal{G}$. Under the thin accretion disk model, a larger observer inclination $\theta_o$ deforms the inner shadow into a hat-like structure. The parameters $(a, Q, \mathcal{G})$ mainly affect the size of the inner shadow and the brightness of the critical curve, where increasing these parameters reduces the shadow size and enhances the distinguishability of the critical curve. In the retrograde accretion disk case, the gravitational redshift significantly reduces the observed brightness of the image. In addition, we compute the distribution of the redshift factor on the projection screen. The results indicate that the Doppler effect induced by large $\theta_o$ enhances the blueshift in the image, while the light emitted by particles plunging into the black hole leads to strong redshift near the inner shadow. Finally, we study the polarization images under synchrotron radiation and find that the polarization intensity $\mathcal{P}_o$ reaches its maximum around the lensed image and higher-order images, whereas no polarization vectors appear within the inner shadow. This stands in sharp contrast to horizonless compact objects. These findings contribute to a deeper understanding of the shadow properties of charged black holes within Lorentz-violating gravity.
}}

\thispagestyle{empty}
\newpage
\setcounter{page}{1}


\section{Introduction}
Since its inception, general relativity (GR) has remained a cornerstone of modern physics, offering a mathematically rigorous framework for describing gravitation. With ongoing advancements in observational techniques, the validity of GR has been confirmed by high-precision and high-resolution experiments across a wide range of regimes\cite{abbott2020gw190412,psaltis2020gravitational}, including both weak\cite{will2014confrontation} and strong gravitational fields\cite{akiyama2019first1,akiyama2019first2,akiyama2022first}. These observations not only reinforce the theoretical foundations of GR but also motivate its modification and extension, paving the way for the exploration of a broader class of gravitational theories. In recent years, certain extensions of GR have gained increasing attention due to their potential to address fundamental problems in astrophysics and cosmology\cite{khan2023charged,khan2024circular,mustafa2024imprints,rehman2023matter,mustafa2020realistic}. One such extension involves modifying the Einstein action by introducing the Kalb–Ramond (KR) field\cite{kalb1974classical}, a 2-form field with self-interacting properties. The KR modification is potentially related to heterotic string theory and may be interpreted as a manifestation of closed string excitations\cite{liang2023new}. Due to the nonminimal coupling between the KR field and the Ricci scalar, this model may lead to the violation of Lorentz symmetry\cite{altschul2010lorentz}. Moreover, the KR field possesses theoretical features such as forming a third-rank antisymmetric tensor and serving as a possible source of spacetime torsion\cite{majumdar1999parity,letelier1995spinning}. Further discussions on the role of KR fields in gravitation and particle physics can be found in \cite{kumar2020gravitational,seifert2010monopole,cox2016stability}. Given the close resemblance between the KR field and spacetime torsion, Einstein theory with the KR field as a source is formally equivalent to a class of modified gravity models that incorporate spacetime torsion.

In 1919, a significant milestone in gravitational theory was achieved when the deflection of starlight was observed during a total solar eclipse, providing the first experimental confirmation of GR. Since then, substantial progress has been made in understanding the gravitational deflection of light, a phenomenon now commonly referred to as gravitational lensing\cite{schneider1992gravitational,blandford1992cosmological,petters2012singularity,perlick2004gravitational,congdon2018principles}. Typical gravitational lensing effects arise in weak gravitational fields, where the deflection angles are small and can be described using linear approximations\cite{einstein1917special}. In recent years, advances in observational techniques have enabled the detection of phenomena in the strong deflection regime. Due to their extremely strong gravitational fields, black holes prevent any light from escaping once it crosses the event horizon, rendering them undetectable through direct observation. Consequently, studying the behavior of null geodesics in the vicinity of black holes becomes essential. The analysis of emitted radiation provides critical information about black holes, and this approach becomes especially effective in the presence of strong gravitational lensing. The term “gravitational lens” refers not only to the bending of light trajectories but also to the variation in the apparent brightness of a source induced by gravitational fields. Under the weak-field approximation, lensing effects generally offer an adequate description of observed phenomena. However, luminous rings associated with compact objects such as black holes and boson stars are more likely the result of strong gravitational lensing\cite{cunha2015shadows}. Additionally, studies of accretion disks around Reissner–Nordström (RN) black holes coupled to nonlinear electrodynamic fields\cite{abbas2023accretion}, as well as investigations into the thermal properties and Joule–Thomson expansion behavior of ModMax black holes\cite{shahzad2024thermal}, have provided important insights into black hole dynamics.

Photons falling into the event horizon cannot reach the observer, so the black hole shadow refers to the two-dimensional dark region on the celestial sphere caused by the strong gravitational field of the black hole. Its size and shape are primarily determined by the geometry of the black hole spacetime and the observer's inclination. Careful analysis of the black hole shadow can provide important information about the black hole spacetime\cite{liu2020shadow}. The study of shadows of compact objects, especially black holes, has a long history and has now reached a relatively mature stage. In recent years, extensive attention has been devoted to a variety of accretion models, including spherical accretion models\cite{narayan2019shadow,zeng2020shadows,heydari2023shadows}, both optically and geometrically thin disk models\cite{yang2024shadow,he2024observational,zeng2022shadows,li2021shadows}, and geometrically thick disk models\cite{vincent2011gyoto,zhang2024images,zhang2024imaging,gjorgjieski2024comparison}. The Event Horizon Telescope (EHT) observations marked a new era in black hole shadow studies. Numerous studies have compared theoretical predictions with observed black hole shadows\cite{zeng2025holographic,cui2024optical,hou2024unique,huang2024images} and constrained various black hole model parameters to ensure consistency between theoretical shadow sizes and those observed for M87* and Sgr A*. In addition, studies of optical images of boson stars have attracted growing interest. A considerable amount of work has been devoted to distinguishing the optical images of boson stars from black hole shadows\cite{maso2021boson,rosa2023imaging,rosa2024accretion,zeng2025opticalimagesminiboson,herdeiro2021imitation,rosa2022shadows,he2025optical}.

At present, studies on the properties of black holes in KR gravity have made systematic and comprehensive progress\cite{zahid2024shadow,duan2024electrically}. In the context of black hole shadows, Kumar et al. analyzed the shadow profile and strong gravitational lensing effects of rotating black holes in KR gravity\cite{kumar2020gravitational}. Zubair et al. further investigated the shadow features of rotating black holes under specific parameter conditions and compared them with the EHT observational data\cite{zubair2023rotating}. Liu et al. presented the shadow images of slowly rotating KR black holes with spherical light sources\cite{liu2024shadow}. However, these studies mainly focused on the slowly rotating case or did not consider the influence of accretion disk models. In our previous work, we analyzed the shadow images of black holes with general rotation parameters in KR gravity under different light source models. The results showed that the observer inclination significantly affects the shape of the inner shadow and the brightness distribution of the higher-order images, while the rotation parameter and the Lorentz-violating parameter primarily influence the size of the inner shadow. Nevertheless, the effect of electric charge on black hole shadows has not yet been systematically investigated in the existing literature. It is well known that, due to the presence of charge, charged black holes exhibit a series of distinctive features in both geometric structure and thermal properties. As a continuation of our previous study, the present work aims to investigate the shadow images under thin accretion disks and their polarization images, and to further explore the influence of Lorentz-violating gravity.

The structure of this paper is as follows. Section~\ref{sec2} briefly reviews the properties of rotating charged black holes in KR gravity and derives the null geodesic equations. Based on these results, we calculate the angular diameter of the shadow and compare it with EHT observational data to constrain the black hole parameters. In Section~\ref{sec3}, we introduce thin accretion disk sources and present shadow images of black holes on the observation plane of the ZAMO. Section~\ref{sec4} analyzes the influence of black hole parameters on the distribution of the redshift factor. Section~\ref{sec5} discusses polarization images under synchrotron radiation. Finally, Section~\ref{sec6} summarizes and discusses the main findings of this work. Throughout this paper, Greek indices $\alpha, \beta, \mu, \nu$ denote spacetime coordinates $(t, x, y, z)$ or $(t, r, \theta, \varphi)$, while Latin indices $i, j$ represent spatial coordinates $(x, y, z)$ or $(r, \theta, \varphi)$. Unless otherwise specified, we adopt geometric units with $c = G = 1$, where $c$ is the speed of light in vacuum and $G$ is the gravitational constant.

\section{Rotating Charged Black Holes in Kalb-Ramond Gravity}\label{sec2}

We begin by briefly introducing the rotating charged black holes in KR gravity, which form the foundation of the present study. The spacetime geometry surrounding a static, spherically symmetric black hole can be described in spherical coordinates as
\begin{equation}
	ds^{2} = -\mathcal{A}(r)dt^{2} + \frac{1}{\mathcal{A}(r)}dr^{2} + r^{2}\left(d\theta^{2} + \sin^{2}\theta d\varphi^{2}\right). \label{eq:nrm}
\end{equation}
In this work, we consider the following form of the radial function $\mathcal{A}(r)$~\cite{duan2024electrically}
\begin{equation}
	\mathcal{A}(r) = \frac{1}{1 - \mathcal{G}} - \frac{2M}{r} + \frac{Q^{2}}{r^{2}(1 - \mathcal{G})^{2}},
\end{equation}
where $M$ and $Q$ denote the mass and charge of the black hole, respectively. The Lorentz-violating parameter $\mathcal{G}$ is a dimensionless quantity that characterizes the effect of Lorentz symmetry violation induced by a nonzero vacuum expectation value of the KR field. According to classical gravitational experiments within the Solar System, $\mathcal{G}$ is tightly constrained and can only take very small values~\cite{yang2023static}. When $\mathcal{G} = 0$, the metric~(\ref{eq:nrm}) reduces to the Reissner–Nordström (RN) metric; when both $\mathcal{G} = 0$ and $Q = 0$, it further reduces to the Schwarzschild metric. The event horizon of a non-rotating charged KR black hole exists only when the condition $Q^2/M^2 \leq (1 - \mathcal{G})^3$ is satisfied, with equality corresponding to the extremal case. Compared to the RN black hole, a positive $\mathcal{G}$ allows an extremal KR black hole to require a smaller charge $Q$, while a negative $\mathcal{G}$ demands a larger $Q$. In addition, the Kretschmann scalar for this spacetime is given by
\begin{equation}
	K = R^{\mu\nu\lambda\sigma} R_{\mu\nu\lambda\sigma} = \frac{48M^2}{r^6} - \frac{16\mathcal{G}M}{(1 - \mathcal{G})r^5} + \frac{4\mathcal{G}^2}{(1 - \mathcal{G})^2r^4} + \frac{56Q^4}{(1 - \mathcal{G})^4r^8} - \frac{96MQ^2}{(1 - \mathcal{G})^2r^7} + \frac{8\mathcal{G}Q^2}{(1 - \mathcal{G})^3r^6}.
\end{equation}
As $r \to 0$, $K \to \infty$, indicating the presence of a scalar polynomial (s.p.) curvature singularity that cannot be removed by any coordinate transformation and originates from the Lorentz-violating effect. It is also worth noting that as $r \to \infty$, $\mathcal{A}(r) \to 1/(1 - \mathcal{G})$, and not all components of the Riemann tensor vanish. This implies that the spacetime is not asymptotically Minkowski.

As a natural generalization of the static case, we now introduce the rotating charged black hole in KR gravity. Since supermassive black holes are generally expected to possess spin, and observational data from the EHT indicate that both M87* and Sgr A* exhibit rotation, the study of rotating black holes is of significant physical interest. The presence of an additional spin parameter introduces substantial differences in the dynamical behavior of rotating black holes compared to their static counterparts, resulting in distinct null geodesics and shadow structures. The Newman–Janis algorithm can be employed to generate the rotating solution from the static metric. The detailed derivation can be found in~\cite{zahid2024shadow} and is not repeated here. The line element of the rotating charged black hole in KR gravity takes the form
\begin{align}
	ds^{2} &= -\left(\frac{\Delta - a^{2} \sin^{2}\theta}{\Sigma}\right) dt^{2} + \frac{\Sigma}{\Delta} dr^{2} + \Sigma d\theta^{2} \nonumber \\
	&\quad + \frac{\sin^{2}\theta}{\Sigma} \left[(r^{2} + a^{2})^{2} - \Delta a^{2} \sin^{2}\theta \right] d\varphi^{2} + \frac{2a \sin^{2}\theta}{\Sigma} \left[\Delta - (r^{2} + a^{2})\right] dt d\varphi, \label{eq:rm}
\end{align}
where
\begin{align}
	\Delta &= a^{2} - 2Mr + \frac{1}{1 - \mathcal{G}} \left(r^{2} + \frac{Q^{2}}{1 - \mathcal{G}}\right), \label{delta} \\
	\Sigma &= a^{2} \cos^{2}\theta + r^{2}.
\end{align}
Here, $a$ is the rotation parameter. The position of the event horizon is determined by solving $\Delta = 0$, yielding
\begin{equation}
	r_{\pm} = M(1 - \mathcal{G}) \pm \sqrt{M^2(1 - \mathcal{G})^2 - \frac{Q^2}{1 - \mathcal{G}} - a^2(1 - \mathcal{G})}.
\end{equation}
The “$+$” sign corresponds to the event horizon, while the “$-$” sign corresponds to the Cauchy horizon. Clearly, the horizon structure differs significantly from that of the Kerr black hole and explicitly depends on the Lorentz-violating parameter $\mathcal{G}$. The sign and magnitude of $\mathcal{G}$ affect both the number and location of the horizons, which in turn influence the motion of particles and the morphology of the black hole shadow. It is evident that the metric~(\ref{eq:rm}) is independent of both $t$ and $\varphi$, i.e., $\partial g_{\mu\nu}/\partial t = \partial g_{\mu\nu}/\partial \varphi = 0$, indicating that the spacetime is stationary and axisymmetric, and admits two Killing vector fields $(\partial/\partial t)^\mu$ and $(\partial/\partial \varphi)^\mu$. When $\mathcal{G} = 0$, the metric~(\ref{eq:rm}) reduces to that of the Kerr–Newman black hole. In this work, our analysis will focus on the rotating metric~(\ref{eq:rm}). Due to the nonlinear terms in Eq.~(\ref{delta}), we solve the equations numerically and fix the black hole mass as $M = 1$ throughout the paper.

To obtain the shadow of a black hole, one must study the null geodesic equations, which can be derived using the Hamilton–Jacobi formalism~\cite{carter1968global}. The primary approaches for analyzing the motion of particles and objects are the Lagrangian and Hamiltonian formalisms. In GR, aside from the mass of the moving particle, these methods usually yield only two conserved quantities during motion: the energy $E$ and the angular momentum $L$. To render the equations fully integrable, an additional conserved quantity is required. The Hamilton–Jacobi formalism allows the $r$ and $\theta$ coordinates in the Hamilton–Jacobi equation to be separated, leading to an extra constant of motion known as the Carter constant~\cite{carter1968global,chandrasekhar1998mathematical}.

In this work, we adopt the Lagrangian formalism, specifically using the Lagrangian defined as
\begin{equation}
	\mathcal{L}(q,\dot{q}) = \frac{1}{2} g_{\mu\nu} \dot{q}^\mu \dot{q}^\nu, \label{eq:la}
\end{equation}
where the dot “$\cdot$” denotes differentiation with respect to the affine parameter $\lambda$ along the null geodesics. The metric $g_{\mu\nu}$ is the same as that given in Eq.~(\ref{eq:rm}). By expanding the Lagrangian~(\ref{eq:la}), two conserved quantities along the geodesics can be obtained
\begin{align}
	E &= \frac{\partial \mathcal{L}}{\partial \dot{t}} = -g_{tt} \dot{t} - g_{\varphi t} \dot{\varphi} = p_t, \\
	L &= \frac{\partial \mathcal{L}}{\partial \dot{\varphi}} = g_{\varphi t} \dot{t} + g_{\varphi\varphi} \dot{\varphi} = p_\varphi.
\end{align}
The Hamilton–Jacobi equation is given by
\begin{equation}
	2 \frac{\partial \mathcal{J}}{\partial \lambda} + g^{\mu\nu} \frac{\partial \mathcal{J}}{\partial x^\mu} \frac{\partial \mathcal{J}}{\partial x^\nu} = 0,
\end{equation}
where
\begin{equation}
	\mathcal{J} = \frac{1}{2} m_0 \lambda - E t + L \varphi + \mathcal{J}_r + \mathcal{J}_\theta
\end{equation}
is the Jacobi action, with $\mathcal{J}_r$ and $\mathcal{J}_\theta$ being functions of $r$ and $\theta$, respectively. Here $m_0$ denotes the mass of a particle moving around the black hole. For photons, $m_0 = 0$. Applying separation of variables to the Hamilton–Jacobi action yields the null geodesic equations
\begin{align}
	&\Sigma \dot{t} = a(L - aE \sin^2 \theta) + \frac{r^2 + a^2}{\Delta} \left[E(r^2 + a^2) - aL \right], \label{geo1} \\
	&\Sigma^2 \dot{r}^2 = R(r), \\
	&\Sigma^2 \dot{\theta}^2 = \Theta(\theta), \label{dottheta} \\
	&\Sigma \dot{\varphi} = \left(L \csc^2 \theta - aE\right) - \frac{a}{\Delta} \left[aL - E(r^2 + a^2) \right], \label{geo4}
\end{align}
where
\begin{align}
	&R(r) = \left[E(r^2 + a^2) - aL \right]^2 - \Delta \left[\mathcal{Q} + (L - aE)^2 \right], \\
	&\Theta(\theta) = \mathcal{Q} + \cos^2 \theta \left(a^2 E^2 - L^2 \csc^2 \theta \right),
\end{align}
are referred to as the radial and angular potentials, respectively, and $\mathcal{Q}$ is the Carter constant. The photon trajectories near the black hole are accurately described by Eqs.~(\ref{geo1})--(\ref{geo4}).

If the photon trajectory lies in the equatorial plane of the black hole, i.e., $\theta = 90^\circ$, the radial equation can be expressed in terms of the effective potential $V_{\mathrm{eff}}$ as
\begin{equation}
	V_{\mathrm{eff}} = -\frac{\left[E(a^2 + r^2) - aL\right]^2 - \Delta \left[\mathcal{Q} + (L - aE)^2\right]}{r^4}.
\end{equation}
We are particularly interested in the unstable null circular orbits near the event horizon, where the photon remains confined to a 2-sphere of constant radius. This surface is referred to as the photon sphere. Photons that enter the photon sphere eventually fall into the event horizon, whereas only those outside the photon sphere may reach the observer. Therefore, the boundary of the black hole shadow is determined by the radius of the photon sphere, which satisfies the following equations
\begin{equation}
	\frac{\partial r(\lambda)}{\partial \lambda} = 0, \quad \frac{\partial^2 r(\lambda)}{\partial \lambda^2} = 0.
\end{equation}
These conditions can equivalently be written as
\begin{equation}
	R(r) = 0, \quad \frac{\partial R(r)}{\partial r} = 0.
\end{equation}
The condition for a circular orbit is determined by the effective potential and requires
\begin{equation}
	V_{\mathrm{eff}}(r) = 0, \quad \frac{\partial V_{\mathrm{eff}}(r)}{\partial r} = 0. \label{eq:effs}
\end{equation}
In addition, an unstable orbit satisfies the condition $\frac{\partial^2 V_{\mathrm{eff}}(r)}{\partial r^2} < 0$ for a local maximum.

To solve the system of Eq.~(\ref{eq:effs}), two impact parameters are defined as follows~\cite{chandrasekhar1998mathematical}
\begin{equation}
	\Xi \equiv \frac{L}{E}, \quad H \equiv \frac{\mathcal{Q}}{E^{2}}.
\end{equation}
The introduction of these impact parameters imposes strict constraints on photon trajectories to ensure the existence of unstable circular orbits. By solving Eq.~(\ref{eq:effs}) under these definitions, we obtain
\begin{align}
	\Xi &= \frac{-4r\Delta + \Delta^{\prime}(r^2 + a^2)}{a\Delta^{\prime}}, \\
	H &= \frac{r^2}{a^2 \Delta^{\prime 2}} \left[ 16\Delta(a^2 - \Delta) - r^2 \Delta^{\prime 2} + 8r\Delta\Delta^{\prime} \right],
\end{align}
where $\Delta^{\prime}$ denotes the derivative of $\Delta$ with respect to the radial coordinate $r$. It is well known that the shadow image of a black hole is always two-dimensional and can be projected onto the Cartesian celestial plane. The coordinates $\alpha$ and $\beta$ on this plane are given by
\begin{align}
	\alpha(\theta \to \theta_o) &= -\lim_{r \to \infty} \left(r^2 \sin \theta \frac{d\varphi}{dr} \right), \label{eq:ccpc1} \\
	\beta(\theta \to \theta_o) &= \lim_{r \to \infty} \left(r^2 \frac{d\theta}{dr} \right), \label{eq:ccpc2}
\end{align}
where $\theta_o$ denotes the inclination angle of the observer and the limit $r \to \infty$ corresponds to the observer being located at spatial infinity. Solving Eqs.~(\ref{eq:ccpc1}) and~(\ref{eq:ccpc2}) and taking the limit yields
\begin{align}
	\alpha &= -H \csc \theta_o, \\
	\beta^2 &= a^2 \cos^2 \theta_o + \Xi - H^2 \cot^2 \theta_o.
\end{align}
The black hole shadow contour can then be plotted in terms of the celestial coordinates $\alpha$ and $\beta$. The effects of different parameters and the observer inclination on the shadow have been discussed in~\cite{zahid2024shadow}. In the following, we attempt to constrain the black hole parameters using EHT observational data.

To effectively characterize the size of the black hole shadow, one can introduce an observable quantity, the radius, defined as
\begin{equation}
	\mathcal{R}_{\mathrm{BH}} = \frac{(\alpha_t - \alpha_r)^2 + \beta_t^2}{2|\alpha_t - \alpha_r|},
\end{equation}
where $\mathcal{R}_{\mathrm{BH}}$ denotes the radius of a reference circle passing through the top, bottom, and rightmost points of the black hole shadow, used to approximate its overall size. Specifically, $\alpha$ and $\beta$ represent the horizontal and vertical celestial coordinates, respectively, and the subscripts $t$ and $r$ refer to the top and right boundaries of the shadow. In astronomical observations, the angular diameter is used to quantify the size of the shadow and is given by
\begin{equation}
	\Theta_{\mathrm{BH}} = 2 \tilde{\mathcal{R}}_{\mathrm{BH}} \frac{\mathcal{M}}{D_o},
\end{equation}
where $D_o$ is the distance between the black hole and the observer, $\tilde{\mathcal{R}}_{\mathrm{BH}}$ is the shadow radius projected onto a screen placed at the black hole's location, and $\mathcal{M}$ is the black hole mass. The radius $\tilde{\mathcal{R}}_{\mathrm{BH}}$ is related to the size $\mathcal{R}_{\mathrm{BH}}$ through a simple geometric relation. As described in~\cite{amarilla2012shadow,li2024shadow}, when the black hole is far from the observer, the angular diameter can be expressed as
\begin{equation}
	\Theta_{\mathrm{BH}} = 2 \times 9.87098, \tilde{\mathcal{R}}_{d} \left( \frac{\mathcal{M}}{M_{\odot}} \right) \left( \frac{1\ \mathrm{kpc}}{D_o} \right)\ \mu \mathrm{as}.
\end{equation}
Using the above formula, we assume that the background spacetime of Sgr A* and M87* is described by Eq.~(\ref{eq:rm}). For Sgr A*, the distance to Earth is $D_o = 8\ \mathrm{kpc}$, and the estimated mass is $\mathcal{M} = \left(4.0_{-0.6}^{+1.1}\right) \times 10^6 M_{\odot}$. The angular diameter observed by the EHT is $\Theta_{\mathrm{Sgr A^*}} = \left(48.7 \pm 7\right)\ \mu \mathrm{as}$~\cite{walia2022testing}. For M87*, the distance is $D_o = 16.8\ \mathrm{kpc}$, and the mass is estimated as $\mathcal{M} = (6.5 \pm 0.7) \times 10^6 M_{\odot}$. The observed angular diameter is $\Theta_{\mathrm{M87^*}} = (37.8 \pm 2.7)\ \mu \mathrm{as}$, showing approximately a $10\%$ deviation from the theoretical prediction~\cite{capozziello2023testing}.

\begin{figure}[!h]
	\centering 
	\subfigure[$a=0.5,\mathcal{G}=0.15$]{\includegraphics[scale=0.7]{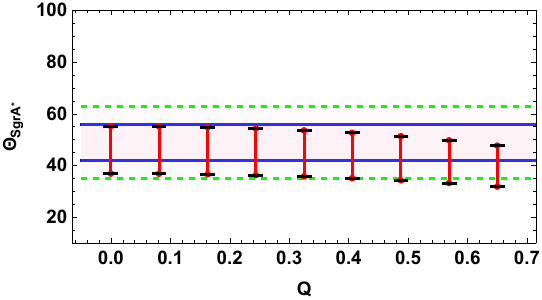}}
	\subfigure[$Q=0.1,a=0.5$]{\includegraphics[scale=0.7]{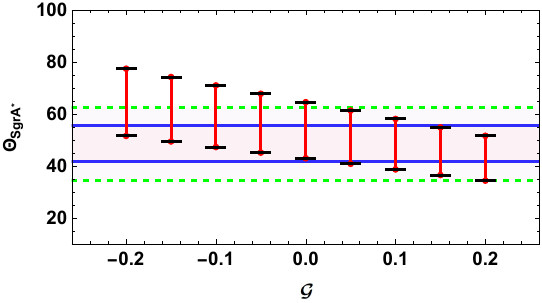}}
	\subfigure[$a=0.5,\mathcal{G}=0.15$]{\includegraphics[scale=0.7]{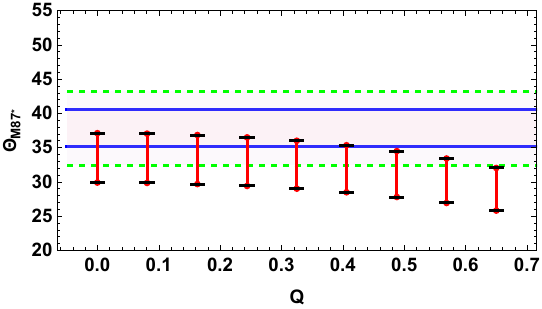}}
	\subfigure[$Q=0.1,a=0.5$]{\includegraphics[scale=0.7]{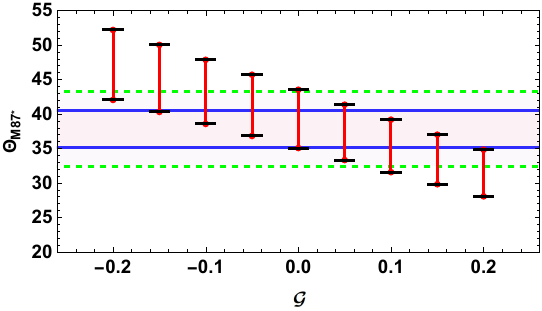}}
	
	\caption{\label{fig1}Estimated ranges of the angular diameter $\Theta_{\mathrm{BH}}$ of the black hole shadow for Sgr A* and M87*. The solid purple lines and dashed green lines represent the $1\sigma$ and $2\sigma$ confidence intervals respectively, and the red segments indicate the estimated ranges.}
\end{figure}
As shown in Figure~\ref{fig1}, we present the estimated ranges of the angular diameter $\Theta_{\mathrm{BH}}$ for the shadows of Sgr A* and M87*. Since the rotation parameter $a$ mainly affects the deviation from circularity rather than the shadow radius, we focus on the effects of the charge $Q$ and the Lorentz-violating parameter $\mathcal{G}$ on the estimated intervals. The first row of Figure~\ref{fig1} corresponds to $\Theta_{\mathrm{Sgr A^*}}$ and the second row to $\Theta_{\mathrm{M87^*}}$. The results show that for $a = 0.5$ and $\mathcal{G} = 0.15$, increasing $Q$ leads to a decreasing trend in the estimated ranges of both $\Theta_{\mathrm{Sgr A^*}}$ and $\Theta_{\mathrm{M87^*}}$. Specifically, for Sgr A*, the range of $\Theta_{\mathrm{Sgr A^*}}$ changes slowly when $Q < 0.25$ but decreases rapidly when $Q > 0.25$. For M87*, this threshold occurs around $Q = 0.3$. Notably, when $Q \approx 0.4$ and $Q \approx 0.65$, the estimated interval of $\Theta_{\mathrm{M87^*}}$ exceeds the $1\sigma$ and $2\sigma$ confidence intervals, respectively. In contrast, for Sgr A*, even as $Q$ increases from $0$ to $0.65$, all corresponding ranges remain within the $1\sigma$ confidence interval. Unlike $Q$, the Lorentz-violating parameter $\mathcal{G}$ exerts a more significant influence on the estimation of $\Theta_{\mathrm{BH}}$. For both $\Theta_{\mathrm{Sgr A^*}}$ and $\Theta_{\mathrm{M87^*}}$, the estimated ranges decrease markedly with increasing $\mathcal{G}$. For Sgr A*, the range of $\Theta_{\mathrm{Sgr A^*}}$ remains within the $1\sigma$ interval when $\mathcal{G} \in (-0.2, 0.2)$. However, for M87*, this condition holds only when $\mathcal{G} \in (-0.15, 0.2)$. These findings suggest that, compared to Sgr A*, the observational data of M87* impose stronger constraints on the parameters $Q$ and $\mathcal{G}$ of rotating charged black holes in KR gravity.

\section{Shadows under Thin Accretion Disks}\label{sec3}

In this section, we explore the shadow images of black holes illuminated by thin accretion disks. As mentioned earlier, the motion of photons within the photon sphere is governed by the impact parameters $\Xi$ and $H$, which determine the boundary of the black hole shadow. Based on this framework, we employ the backward ray-tracing technique \cite{cunha2015shadows} to simulate the shadow. The key to imaging is to distinguish between light rays that reach the observer and those that are captured by the event horizon, with the latter delineating the dark regions on the projection screen. 

\subsection{Image Projection in the ZAMO Frame}

We first establish a local basis coordinate system at a radial distance $r \to r_o$ from the black hole. The coordinates of the observer’s local basis $\{e_{\bar{t}},e_{\bar{r}},e_{\bar{\theta}},e_{\bar{\varphi}}\}$ can be transformed into the coordinate basis of the black hole spacetime $\{\partial_t,\partial_r,\partial_\theta,\partial_\varphi\}$ via the relation
\begin{equation}
	e_{\bar{\mu}} = e_{\bar{\mu}}^{\ \nu} \partial_{\nu},
\end{equation}
where $e_{\bar{\mu}}^{\ \nu}$ is the transformation matrix, which satisfies
\begin{equation}
	g_{\mu\nu} e_{\bar{\alpha}}^{\ \mu} e_{\bar{\beta}}^{\ \nu} = \eta_{\bar{\alpha} \bar{\beta}},
\end{equation}
with $\eta_{\bar{\alpha} \bar{\beta}}$ denoting the Minkowski metric. It is worth noting that the choice of $e_{\bar{\mu}}^{\ \nu}$ is not unique, as different transformation matrices are related by spatial rotations or Lorentz transformations. In this paper, we adopt the specific choice given in \cite{wang2020shadow},
\begin{equation}
	e_{\bar{\mu}}^{\ \nu} = 
	\left(
	\begin{array}{cccc}
		A_{00} & 0 & 0 & A_{03} \\
		0 & A_{11} & 0 & 0 \\
		0 & 0 & A_{22} & 0 \\
		0 & 0 & 0 & A_{33}
	\end{array}
	\right).
\end{equation}
For a zero-angular-momentum observer (ZAMO), i.e., one whose axial angular momentum vanishes at spatial infinity, the components of $e_{\bar{\mu}}^{\ \nu}$ are real and explicitly given by
\begin{align}
	&A_{00} = \sqrt{\frac{g_{\varphi\varphi}}{g_{t\varphi}^2 - g_{tt} g_{\varphi\varphi}}}, \quad 
	A_{03} = -\frac{g_{t\varphi}}{g_{\varphi\varphi}} \sqrt{\frac{g_{\varphi\varphi}}{g_{t\varphi}^2 - g_{tt} g_{\varphi\varphi}}}, \\
	&A_{11} = \frac{1}{\sqrt{g_{rr}}}, \quad
	A_{22} = \frac{1}{\sqrt{g_{\theta\theta}}}, \quad
	A_{33} = \frac{1}{\sqrt{g_{\varphi\varphi}}}.\label{eq:zamo}
\end{align}
Through the above transformation, the photon four-momentum measured in the local inertial frame of the observer can be written as
\begin{equation}
	p^{\bar{t}} = -p_{\bar{t}} = -e_{\bar{t}}^{\ \nu} p_{\nu}, \quad p^{\bar{i}} = p_{\bar{i}} = e_{\bar{i}}^{\ \nu} p_{\nu},
\end{equation}
where $\bar{i} = (r, \theta, \varphi)$ and $\mu, \nu = (t, r, \theta, \varphi)$. In the spacetime described by Eq.~(\ref{eq:rm}), the locally measured components of the photon four-momentum $p^{\bar{\mu}}$ can be explicitly expressed as
\begin{equation}
	p^{\bar{t}} = A_{00} E - A_{03} L, \quad
	p^{\bar{r}} = \frac{p_r}{\sqrt{g_{rr}}}, \quad
	p^{\bar{\theta}} = \frac{p_\theta}{\sqrt{g_{\theta\theta}}}, \quad
	p^{\bar{\varphi}} = \frac{L}{\sqrt{g_{\varphi\varphi}}}, \label{eq:p4m}
\end{equation}
where
\begin{equation}
	p_\theta = g_{\theta\theta} \frac{d\theta}{d\lambda}, \quad
	p_r = g_{rr} \frac{dr}{d\lambda}
\end{equation}
are the covariant momentum components of the photon. In the ZAMO framework, light rays can be projected using celestial coordinates. Let $(\mathcal{X}, \mathcal{Y})$ denote the celestial coordinates. According to \cite{hu2021qed}, the relationship between the photon four-momentum $p_{\bar{\mu}}$ and the celestial coordinates is given by
\begin{equation}
	\cos \mathcal{X} = \frac{p^{\bar{r}}}{p^{\bar{t}}}, \quad
	\tan \mathcal{Y} = \frac{p^{\bar{\varphi}}}{p^{\bar{\theta}}}.
\end{equation}
To obtain the black hole shadow image, the celestial coordinates $(\mathcal{X}, \mathcal{Y})$ must be further projected onto the image plane. On this plane, a standard Cartesian coordinate system $(x, y)$ can be introduced, with the relation
\begin{equation}
	x = -2 \tan\left(\frac{\mathcal{X}}{2}\right) \sin \mathcal{Y}, \quad
	y = -2 \tan\left(\frac{\mathcal{X}}{2}\right) \cos \mathcal{Y}. \label{eq:pro1}
\end{equation}
This projection essentially determines the initial position and momentum of photons in the ZAMO frame. There are various ways to relate the celestial coordinates $(\mathcal{X},\mathcal{Y})$ to the Cartesian coordinates $(x, y)$. In this work, we adopt the fisheye camera model due to its wide field of view. The angle of the field of view $\psi_{\mathrm{fov}}$ determines the extent of the visible region. For simplicity, we set the field of view to $\psi_{\mathrm{fov}}/2$ in both the $x$ and $y$ directions, thereby defining a square screen with side length
\begin{equation}
	L = 2\left|\overrightarrow{V}\right|\tan\left(\frac{\psi_{\mathrm{fov}}}{2}\right),
\end{equation}
where $\overrightarrow{V}$ is the projection of the null geodesic's tangent vector onto the three-dimensional subspace of the observer. The image plane is divided into $N \times N$ pixels, each with a side length of
\begin{equation}
	\ell = \frac{L}{N} = \frac{2}{N}\left|\overrightarrow{V}\right|\tan\left(\frac{\psi_{\mathrm{fov}}}{2}\right).
\end{equation}
We label the pixel centers with coordinates $(u, v)$, where the bottom-left pixel is $(1, 1)$ and the top-right pixel is $(N, N)$, with $u, v = 1, \dots, N$. The relation between the Cartesian coordinates $(x, y)$ and the pixel coordinates $(u, v)$ is given by
\begin{equation}
	x = \ell\left(u - \frac{N+1}{2}\right),\quad 
	y = \ell\left(v - \frac{N+1}{2}\right). \label{eq:pro2}
\end{equation}
Comparing Eqs.~(\ref{eq:pro1}) and (\ref{eq:pro2}), we obtain the transformation between the pixel coordinates $(u, v)$ and the celestial coordinates $(\mathcal{X}, \mathcal{Y})$
\begin{align}
	\mathcal{X} &= 2 \arctan \left[ \frac{1}{N}\tan\left(\frac{\psi_{\mathrm{fov}}}{2}\right) 
	\sqrt{\left(u - \frac{N+1}{2}\right)^2 + \left(v - \frac{N+1}{2}\right)^2}\right], \\
	\mathcal{Y} &= \arctan \left[\frac{2v - (N+1)}{2u - (N+1)}\right].
\end{align}

\subsection{Dynamics of Black Hole Shadows under Thin Accretion Disks}

In realistic astrophysical environments, black holes are often surrounded by accretion disks. Therefore, studying shadow images formed under disk illumination is more physically meaningful than analyzing those produced by spherical light sources. In this work, we assume that the accretion disk lies in the equatorial plane and is both optically and geometrically thin. The observer is positioned at a sufficiently large distance, that is, $r_o \gg M = 1$. Under this condition, the accretion disk can be treated as a collection of electrically neutral plasma moving along timelike geodesics. Material in the accretion disk that approaches the black hole too closely will fall into the event horizon, whereas matter at greater distances can maintain stable circular orbits around the black hole. Therefore, following the approach in~\cite{hou2022image}, we introduce the innermost stable circular orbit (ISCO) as the dividing line. Beyond the ISCO, particles in the disk follow Keplerian orbits, while within the ISCO, they accelerate along critical plunging orbits into the event horizon. This behavior has been confirmed by astronomical observations~\cite{chael2021observing}. 

The location of the ISCO is determined by the effective potential of the particle. For a massive electrically neutral particle moving in the equatorial plane ($\theta=90^\circ$) with four-velocity $u^\mu$, the effective potential is given by
\begin{equation}
	\tilde{V}_{\mathrm{eff}}(r,\tilde{E},\tilde{L}) = 1 + g^{tt} \tilde{E}^2 + g^{\varphi\varphi} \tilde{L}^2 - 2g^{t\varphi} \tilde{E} \tilde{L}.
\end{equation}
Here, $\tilde{E}$ and $\tilde{L}$ denote the specific energy and specific angular momentum of the electrically neutral particle, which are conserved along geodesics. Their expressions are
\begin{align}
	\tilde{E} &= -\frac{g_{tt} + g_{t\varphi} \omega_{\mathrm{e}}}{\sqrt{-g_{tt} - 2g_{t\varphi} \omega_{\mathrm{e}} - g_{\varphi\varphi} \omega_{\mathrm{e}}^2}}, \\
	\tilde{L} &= \frac{g_{t\varphi} + g_{\varphi\varphi} \omega_{\mathrm{e}}}{\sqrt{-g_{tt} - 2g_{t\varphi} \omega_{\mathrm{e}} - g_{\varphi\varphi} \omega_{\mathrm{e}}^2}},
\end{align}
where $\omega_{\mathrm{e}}$ is the angular velocity defined by
\begin{equation}
	\omega_{\mathrm{e}} = \frac{d\varphi}{dt} \equiv \frac{\partial_r g_{t\varphi} + \sqrt{(\partial_r g_{t\varphi})^2 - \partial_r g_{tt} , \partial_r g_{\varphi\varphi}}}{\partial_r g_{\varphi\varphi}}.
\end{equation}
The location $r_I$ of the ISCO must satisfy the following conditions
\begin{equation}
	\left.\tilde{V}_{\mathrm{eff}}\right|_{r = r_I} = 0,\quad \left.\frac{\partial \tilde{V}_{\mathrm{eff}}}{\partial r}\right|_{r = r_I} = 0,\quad \left.\frac{\partial^2 \tilde{V}_{\mathrm{eff}}}{\partial r^2}\right|_{r = r_I} = 0.
\end{equation}
Inside the accretion disk, particles in the region $r_h < r \leq r_I$ plunge into the event horizon along critical orbits. Their four-velocity components satisfy
\begin{align}
	u_\mathrm{in}^t &= -g^{tt} \tilde{E}_i + g^{t\varphi} \tilde{L}_i, \\
	u_\mathrm{in}^r &= -\sqrt{-\frac{1 + g_{tt} (u_\mathrm{in}^t)^2 + 2g_{t\varphi} u_\mathrm{in}^t u_\mathrm{in}^\varphi + g_{\varphi\varphi} (u_\mathrm{in}^\varphi)^2}{g_{rr}}}, \\
	u_\mathrm{in}^\theta &= 0, \\
	u_\mathrm{in}^\varphi &= -g^{t\varphi} \tilde{E}_i + g^{\varphi\varphi} \tilde{L}_i.
\end{align}
Here, $\tilde{E}_i$ and $\tilde{L}_i$ are the conserved quantities evaluated at $r = r_I$, and the negative sign in $u_\mathrm{in}^r$ indicates that the particle is moving toward the black hole. For $r > r_I$, the particles in the accretion disk follow Keplerian orbits, and their four-velocity is given by
\begin{equation}
	u_\mathrm{out}^\mu = \sqrt{\frac{1}{-g_{tt} - 2g_{t\varphi} \omega_{\mathrm{e}} - g_{\varphi\varphi} \omega_{\mathrm{e}}^2}} \left(1, 0, 0, \omega_{\mathrm{e}}\right).
\end{equation}

On the other hand, photons emitted from the accretion disk may cross the black hole’s equatorial plane multiple times before reaching the observer. Each intersection contributes to an increase in image brightness~\cite{hou2024observational}. The image formed by the first crossing is called the direct image, while the second corresponds to the lensed image. When the number of crossings exceeds two, the resulting structures are collectively referred to as higher-order images. Let $r_n$ ($n = 1, 2, \dots, N$) denote the radial coordinates of these intersection points, where $N$ is the maximum number of intersections between the light ray and the disk. The value of $r_n$ determines the shape of the $n$th-order image, with the direct image, lensed image, and higher-order images corresponding to $n = 1$, $n = 2$, and $n > 2$, respectively. In this context, the observed intensity on the screen is given by
\begin{equation}
	I_o = \sum_{n=1}^N \mathcal{F}_n \chi_n^3 \mathcal{E}_n(r), \label{eq:io}
\end{equation}
where $\mathcal{F}_n$ is the fudge factor, $\chi_n$ is the redshift factor, and $\mathcal{E}_n(r)$ is the emissivity. For simplicity, we set $\mathcal{F}_n = 1$. Considering that the black hole images observed by the EHT correspond to a wavelength of $1.3\ \mathrm{mm}$ (230 GHz), the emissivity of the thin disk is modeled as a second-order polynomial in logarithmic space
\begin{equation}
	\mathcal{E}_n(r) = e^{-\frac{1}{2} z^2 - 2z}, \quad z = \log \frac{r}{r_h}.
\end{equation}
The redshift factor is defined as the ratio between the observed frequency and the emission frequency at $r_n$
\begin{equation}
	\chi_n \equiv \frac{\nu_{\mathrm{obs}}}{\nu_{n}},
\end{equation}
where $\nu_{\mathrm{obs}}$ is the frequency received by the observer, and $\nu_{n}$ is the frequency measured in the locally stationary frame co-moving with the emission profiles. The factor $\chi_n^3$ applies to specific frequency intensities, whereas $\chi_n^4$ is used for integrated intensities \cite{wang2024image}. Since the emission spectra of particles differ significantly inside and outside the ISCO, the form of the redshift factor depends on the relation between the radial distance $r$ and $r_I$. For $r > r_I$, the accretion flow moves along circular orbits, and the corresponding redshift factor is given by
\begin{equation}
	\left.\chi_n^\mathrm{out} = \frac{k_2\left(1 - k_1 \frac{p_\varphi}{p_t}\right)}{k_3\left(1 + \omega_{\mathrm{e}} \frac{p_\varphi}{p_t}\right)}\right|_{r = r_n}, \label{rsf1}
\end{equation}
where
\begin{equation}
	k_1 = \frac{g_{t\varphi}}{g_{\varphi\varphi}},\quad 
	k_2 = \sqrt{-\frac{g_{\varphi\varphi}}{g_{tt}g_{\varphi\varphi} - g_{t\varphi}^2}},\quad 
	k_3 = \sqrt{\frac{-1}{g_{tt} + 2g_{t\varphi}\omega_{\mathrm{e}} + g_{\varphi\varphi} \omega_{\mathrm{e}}^2}}.
\end{equation}
In addition, the ratio between the energy observed on the image screen and that propagated along the null geodesics is
\begin{equation}
	d = \frac{p_{t1}}{p_{t2}} = k_2 \left(1 - k_1 \frac{p_\varphi}{p_t}\right).
\end{equation}
For asymptotically flat spacetimes, $d = 1$ when the observer is located at spatial infinity. For $r < r_I$, the accretion flow follows critical plunging orbits, and the corresponding redshift factor is
\begin{equation}
	\left.\chi_n^\mathrm{in} = -\frac{1}{u_\mathrm{in}^{r} p_r/p_t - \tilde{E}_i (g^{tt} - g^{t\varphi} p_\varphi/p_t) + \tilde{E}_i (g^{\varphi\varphi} p_\varphi/p_t + g^{t\varphi})}\right|_{r = r_n}. \label{rsf2}
\end{equation}
Combining the above redshift factors with the radiation model of the thin accretion disk, we can construct shadow images of rotating charged black holes in KR gravity on the image plane.

\subsection{Optical Appearance of Black Holes}
In this subsection, we present the optical appearance of shadow images for rotating charged black holes in KR gravity under thin accretion disks by employing backward ray-tracing techniques. The accretion disk extends from $1.01r_h$ to $60r_h$, where $r_h$ denotes the event horizon of the black hole. The observer is located at a distance of $500M$ from the black hole, with the black hole mass set as $M=1$. The inclination angle $\theta_o$ between the observer and the black hole's rotation axis is taken to be $0^\circ$, $17^\circ$, and $75^\circ$, where $\theta_o = 17^\circ$ corresponds to the viewing angle of M87* as determined by the EHT. Given the intrinsic properties of the spacetime described by (\ref{eq:rm}), we investigate the effects of the rotation parameter $a$, the charge $Q$, and the Lorentz-violating parameter $\mathcal{G}$ on the black hole shadows. On the other hand, the motion of the accretion disk is classified into prograde and retrograde cases. For each set of parameters, we present two types of images. The first shows the intensity distribution, computed from Eq.~(\ref{eq:io}), while the second displays the lensing bands.

\begin{figure}[!h]
	\centering 
	\subfigure[$a=0.1, \theta_o=0^\circ$]{\includegraphics[scale=0.5]{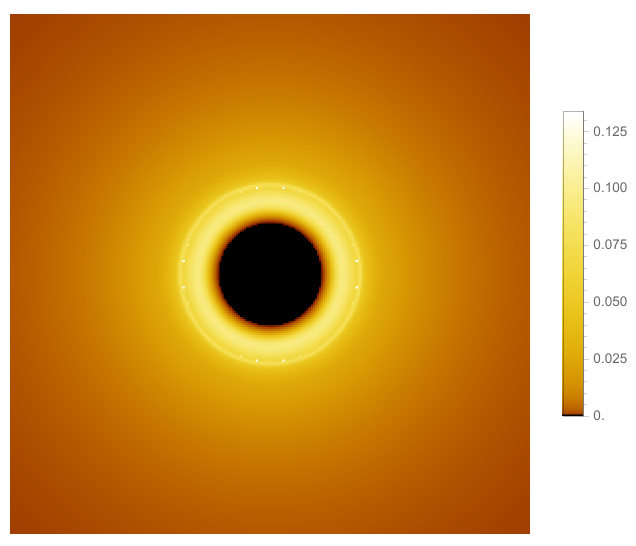}}
	\subfigure[$a=0.1, \theta_o=17^\circ$]{\includegraphics[scale=0.5]{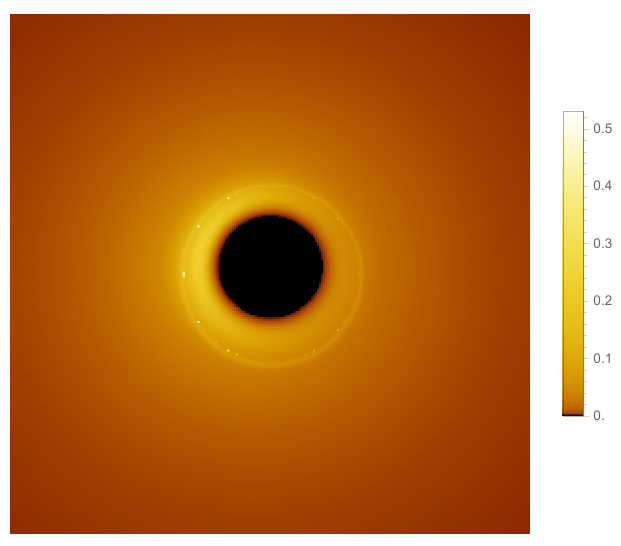}}
	\subfigure[$a=0.1, \theta_o=75^\circ$]{\includegraphics[scale=0.5]{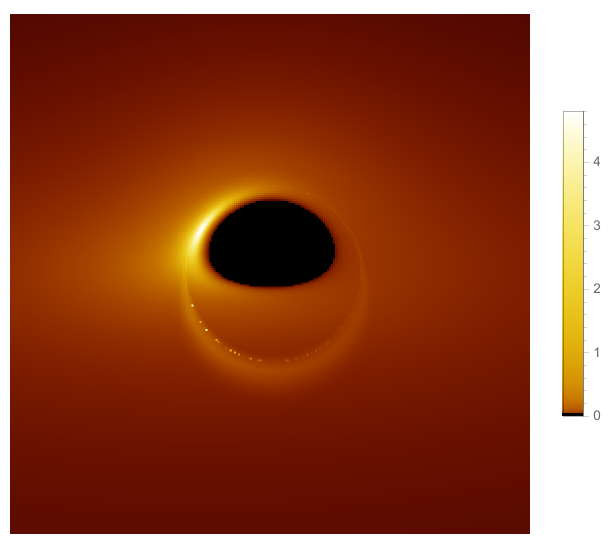}}
	\subfigure[$a=0.5, \theta_o=0^\circ$]{\includegraphics[scale=0.5]{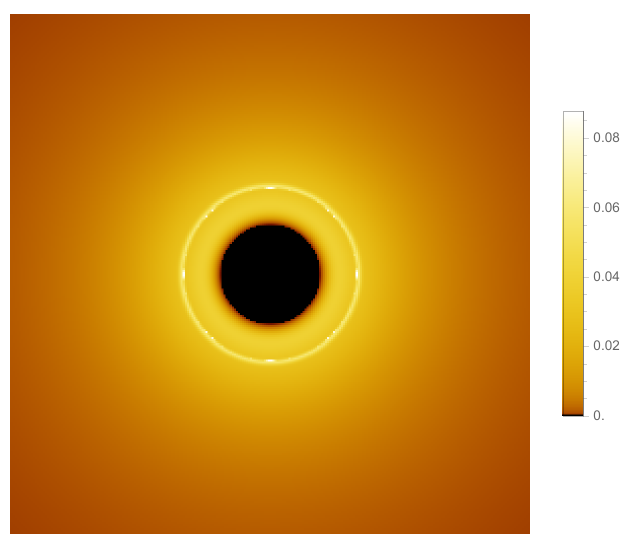}}
	\subfigure[$a=0.5, \theta_o=17^\circ$]{\includegraphics[scale=0.5]{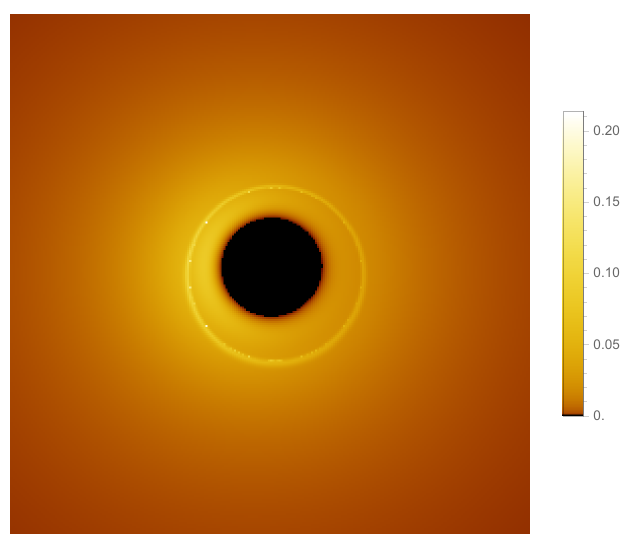}}
	\subfigure[$a=0.5, \theta_o=75^\circ$]{\includegraphics[scale=0.5]{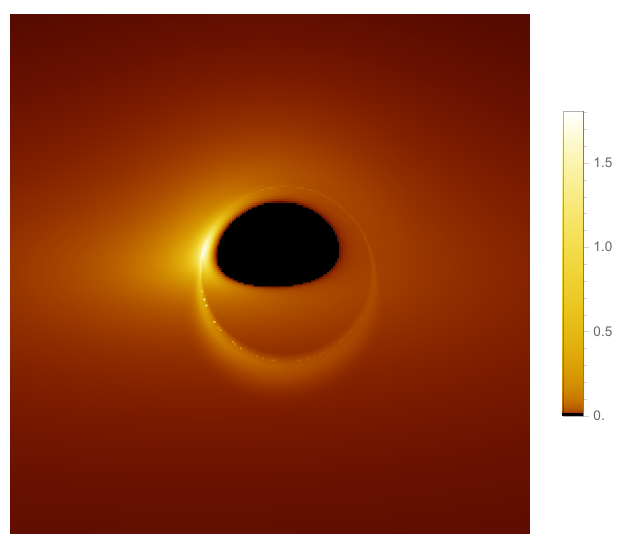}}
	\subfigure[$a=0.9, \theta_o=0^\circ$]{\includegraphics[scale=0.5]{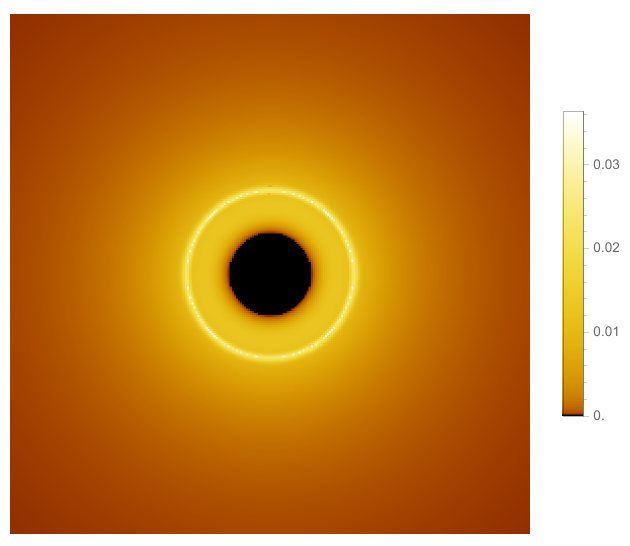}}
	\subfigure[$a=0.9, \theta_o=17^\circ$]{\includegraphics[scale=0.5]{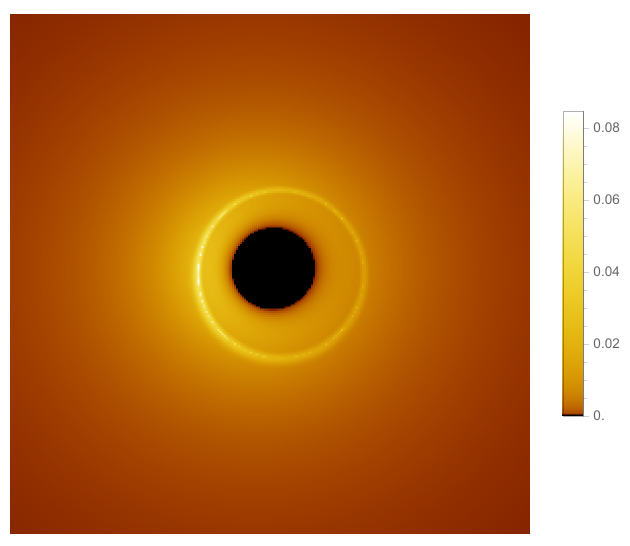}}
	\subfigure[$a=0.9, \theta_o=75^\circ$]{\includegraphics[scale=0.5]{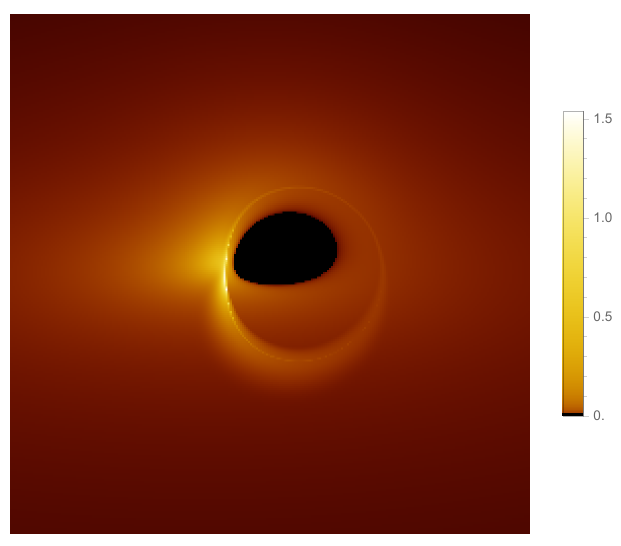}}
	
	\caption{\label{fig3}Shadow images of black holes under a prograde thin accretion disk. The first to third rows correspond to rotation parameter $a = 0.1$, $0.5$, and $0.9$; the first to third columns correspond to observer inclinations $\theta_o = 0^\circ$, $17^\circ$, and $75^\circ$. All other parameters are fixed with charge $Q = 0.1$ and Lorentz-violating parameter $\mathcal{G} = 0.15$.}
\end{figure}

We first focus on the case of a prograde accretion disk. Figure~\ref{fig3} illustrates how the rotation parameter $a$ affects the black hole shadow. It is evident that a central dark region, referred to as the inner shadow~\cite{hou2024observational}, always appears in the images regardless of the values of $a$ and $\theta_o$. This region is formed by photons that fail to reach the observer. When photons emitted from the accretion disk fall directly into the event horizon, corresponding to $n = 0$, (\ref{eq:io}) yields zero intensity, reflecting the fact that the strong gravitational field of the black hole prevents these light rays from escaping. When $\theta_o = 0^\circ$, the inner shadow appears as a perfect circle, whereas at $\theta_o = 75^\circ$, it becomes deformed into a D-shaped silhouette. Another notable feature in the images is a bright, closed curve surrounding the inner shadow, which persists for all parameter configurations. This curve is known as the critical curve. It is worth emphasizing that the presence of both the inner shadow and the critical curve does not depend on the values of $(a, \theta_o)$, indicating that these are intrinsic properties of the spacetime described by (\ref{eq:rm}).

In Figure~\ref{fig3}, as the rotation parameter $a$ increases, the size of the inner shadow decreases significantly, while its shape remains nearly unchanged. In addition, variations in $a$ have a slight effect on the location and size of the critical curve, but a more pronounced impact on its brightness. At low inclination angles ($\theta_o = 0^\circ, 17^\circ$), increasing $a$ leads to a more concentrated distribution of the bright bands associated with the critical curve, thereby enhancing their visual distinguishability. In such cases, the direct and lensed images become difficult to distinguish. At high inclination ($\theta_o = 75^\circ$), a crescent-shaped bright area appears on the left side of the critical curve, exhibiting significantly higher intensity than the surrounding regions. This feature arises from the Doppler effect caused by the relative motion between the prograde accreting material and the observer. Notably, at high inclination, increasing $a$ causes the inner shadow to rotate counterclockwise around the $x$-axis.

\begin{figure}[!h]
	\centering 

	\subfigure[$Q=0$]{\includegraphics[scale=0.35]{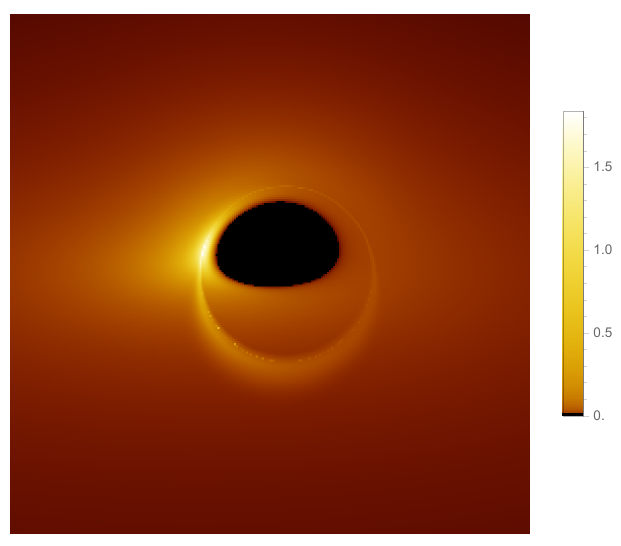}}
	\subfigure[$Q=0.1$]{\includegraphics[scale=0.35]{Thin6_1.pdf}}
	\subfigure[$Q=0.3$]{\includegraphics[scale=0.35]{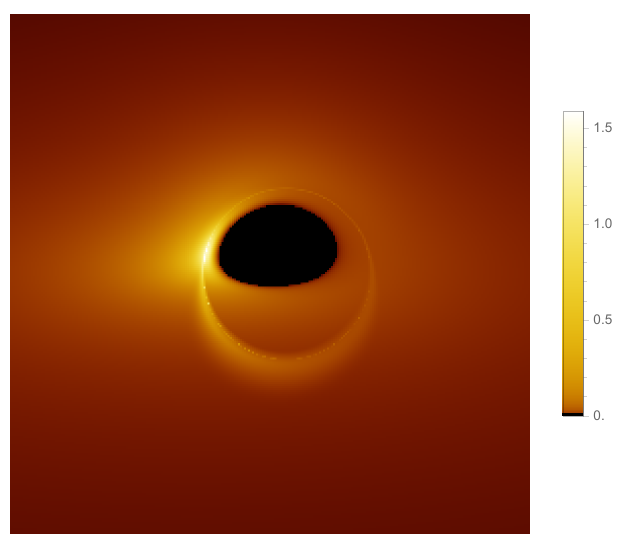}}
	\subfigure[$Q=0.65$]{\includegraphics[scale=0.35]{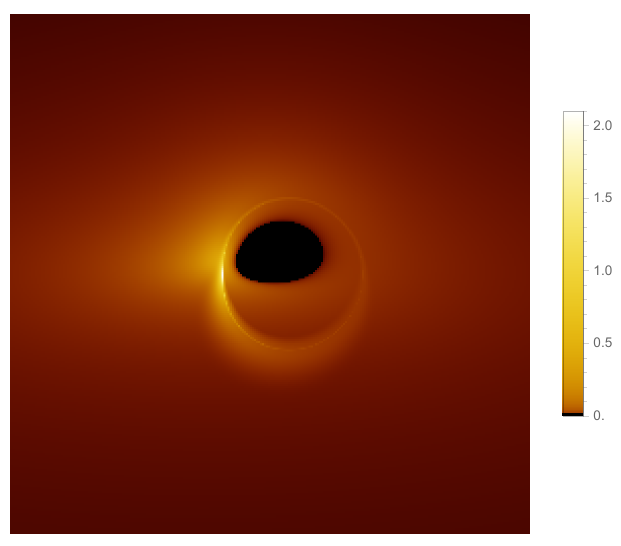}}
	
	\caption{\label{fig4}Shadow images of black holes under a prograde thin accretion disk. From left to right, the charge increases as $Q = 0$, $0.1$, $0.3$, and $0.65$. All other parameters are fixed with the rotation parameter $a = 0.5$, the Lorentz-violating parameter $\mathcal{G} = 0.15$, and the observer inclination $\theta_o = 75^\circ$.}

\end{figure}

\begin{figure}[!h]
	\centering 

	\subfigure[$\mathcal{G}=-0.1$]{\includegraphics[scale=0.35]{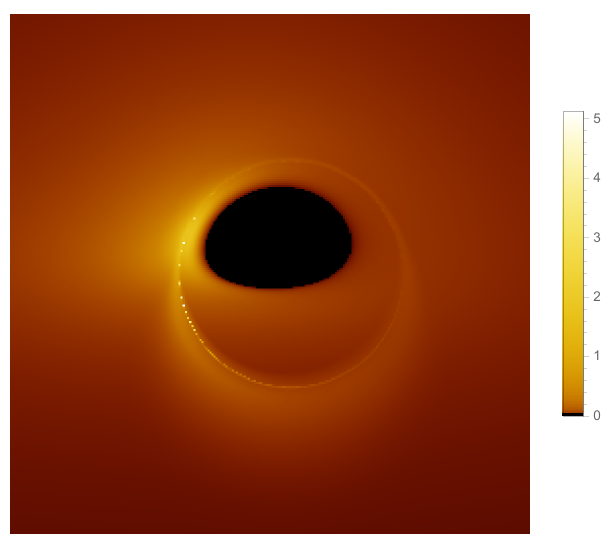}}
	\subfigure[$\mathcal{G}=0$]{\includegraphics[scale=0.35]{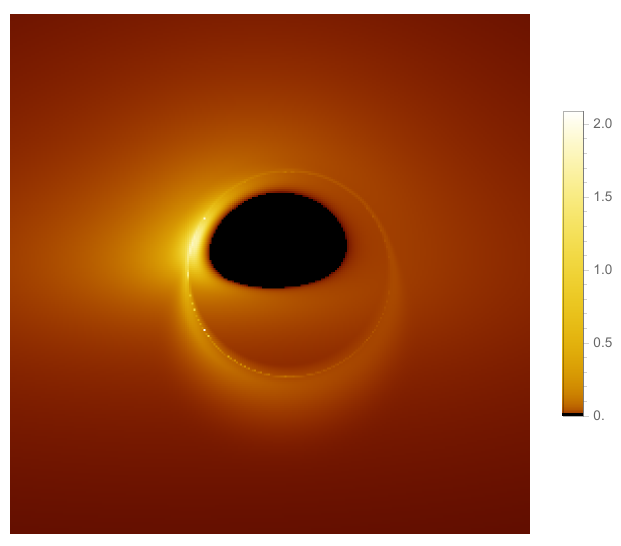}}
	\subfigure[$\mathcal{G}=0.1$]{\includegraphics[scale=0.35]{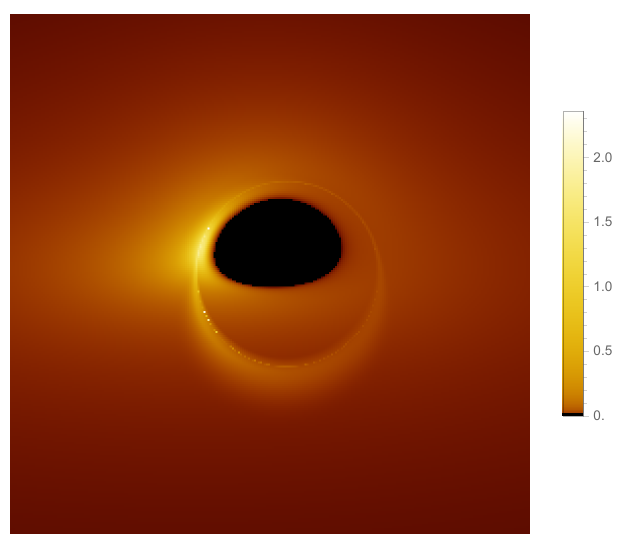}}
	\subfigure[$\mathcal{G}=0.15$]{\includegraphics[scale=0.35]{Thin6_1.pdf}}
	
	\caption{\label{fig5}Shadow images of black holes under a prograde thin accretion disk. From left to right, the Lorentz-violating parameter increases as $\mathcal{G} = -0.1$, $0$, $0.1$, and $0.15$. All other parameters are fixed with the rotation parameter $a = 0.5$, the charge $Q = 0.1$, and the observer inclination $\theta_o = 75^\circ$.}

\end{figure}

Figure~\ref{fig4} illustrates the effect of the charge $Q$ on the black hole shadow. Similar to the influence of $a$, an increase in $Q$ leads to a reduction in the size of the inner shadow while its shape remains unchanged. However, unlike the rotation parameter, $Q$ has a pronounced effect on the size of the critical curve, showing a negative correlation between the two. Moreover, smaller critical curves are observed to be more visually distinguishable. Additionally, due to the large observer inclination angle ($\theta_o = 75^\circ$), a crescent-shaped bright area appears on the left side of the critical curve, as seen in the case of varying $a$. Figure~\ref{fig5} shows the influence of the Lorentz-violating parameter $\mathcal{G}$. It is evident that when $\mathcal{G}$ is positive, the bright band outside the inner shadow becomes more concentrated than in the case of negative $\mathcal{G}$. From left to right, it can be seen that a larger $\mathcal{G}$ leads to a reduction in the size of the inner shadow.

\begin{figure}[!h]
	\centering 
	\subfigure[$a=0.5,Q=0.1,\mathcal{G}=0.15,\theta_o=0^\circ$]{\includegraphics[scale=0.4]{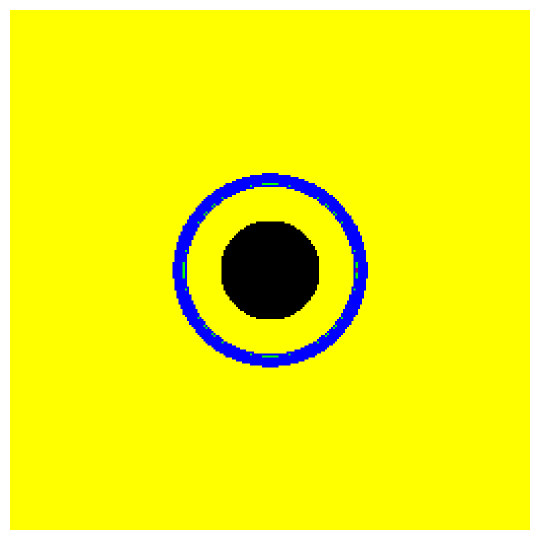}}
	\subfigure[$a=0.9,Q=0.1,\mathcal{G}=0.15,\theta_o=0^\circ$]{\includegraphics[scale=0.4]{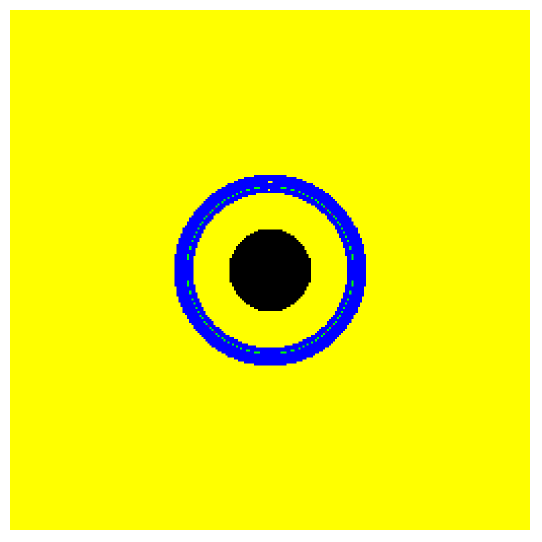}}
	\subfigure[$a=0.5,Q=0.65,\mathcal{G}=0.15,\theta_o=0^\circ$]{\includegraphics[scale=0.4]{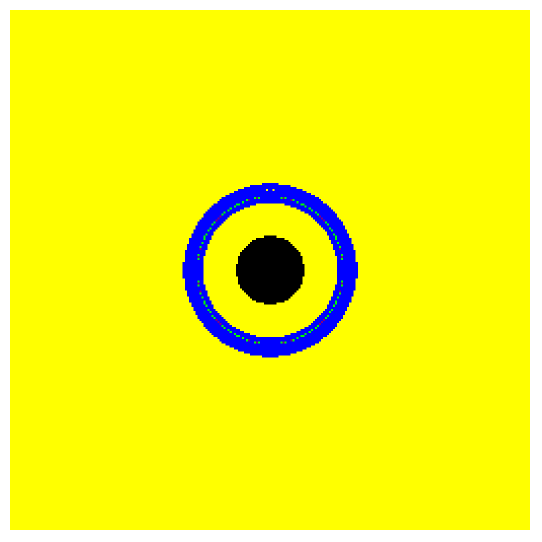}}
	\subfigure[$a=0.5,Q=0.1,\mathcal{G}=-0.1,\theta_o=0^\circ$]{\includegraphics[scale=0.4]{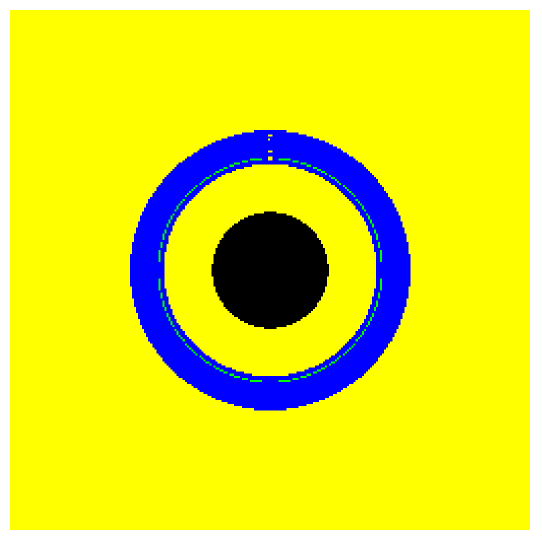}}
	\subfigure[$a=0.5,Q=0.1,\mathcal{G}=0.15,\theta_o=17^\circ$]{\includegraphics[scale=0.4]{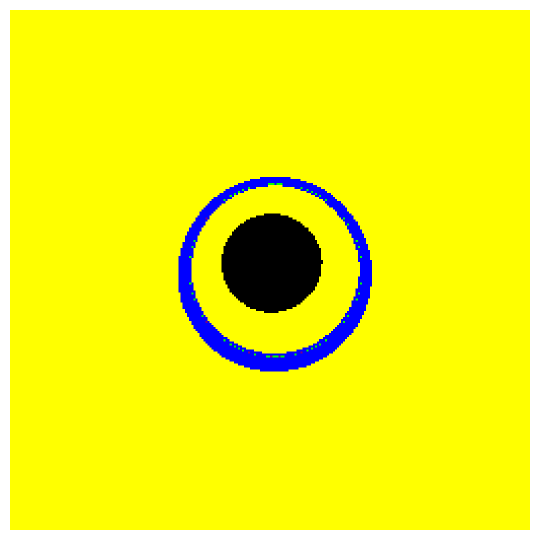}}
	\subfigure[$a=0.9,Q=0.1,\mathcal{G}=0.15,\theta_o=17^\circ$]{\includegraphics[scale=0.4]{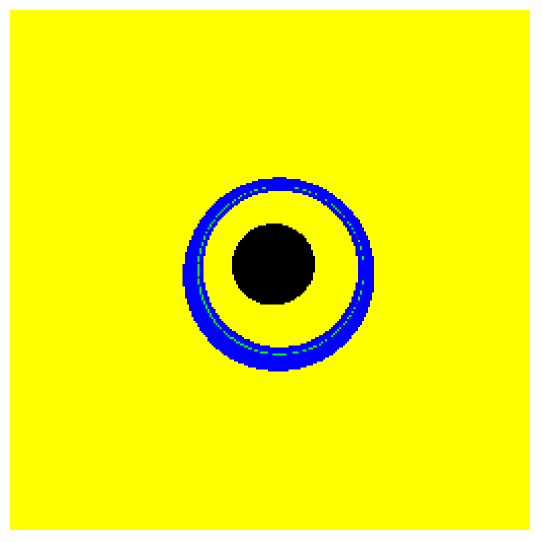}}
	\subfigure[$a=0.5,Q=0.65,\mathcal{G}=0.15,\theta_o=17^\circ$]{\includegraphics[scale=0.4]{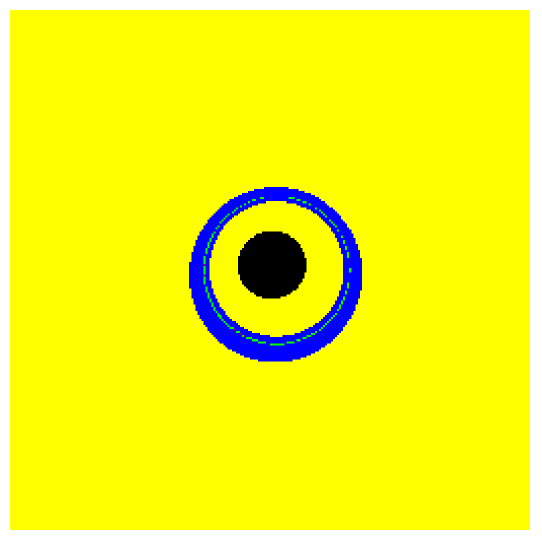}}
	\subfigure[$a=0.5,Q=0.1,\mathcal{G}=-0.1,\theta_o=17^\circ$]{\includegraphics[scale=0.4]{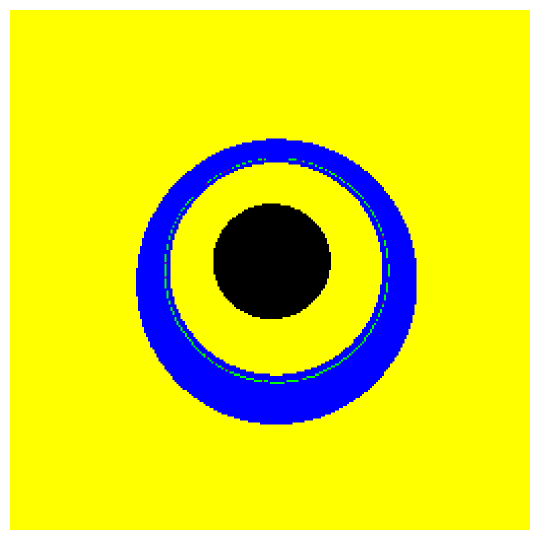}}
	\subfigure[$a=0.5,Q=0.1,\mathcal{G}=0.15,\theta_o=75^\circ$]{\includegraphics[scale=0.4]{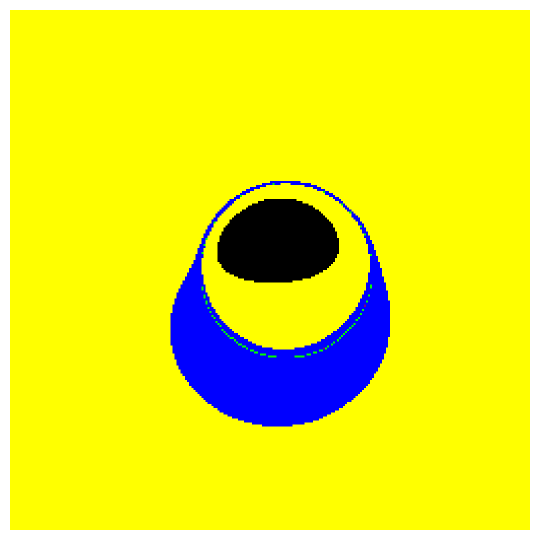}}
	\subfigure[$a=0.9,Q=0.1,\mathcal{G}=0.15,\theta_o=75^\circ$]{\includegraphics[scale=0.4]{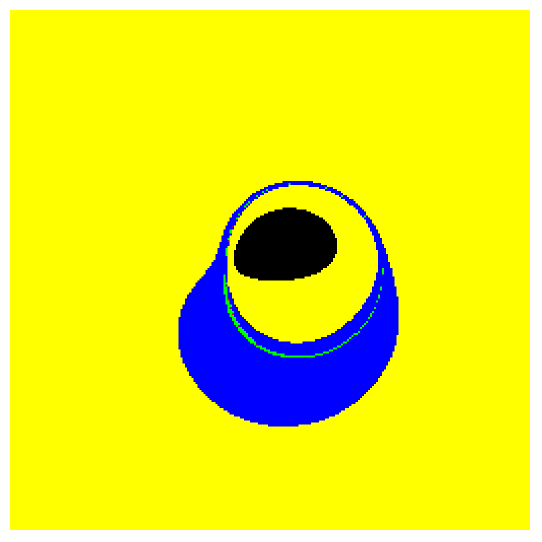}}
	\subfigure[$a=0.5,Q=0.65,\mathcal{G}=0.15,\theta_o=75^\circ$]{\includegraphics[scale=0.4]{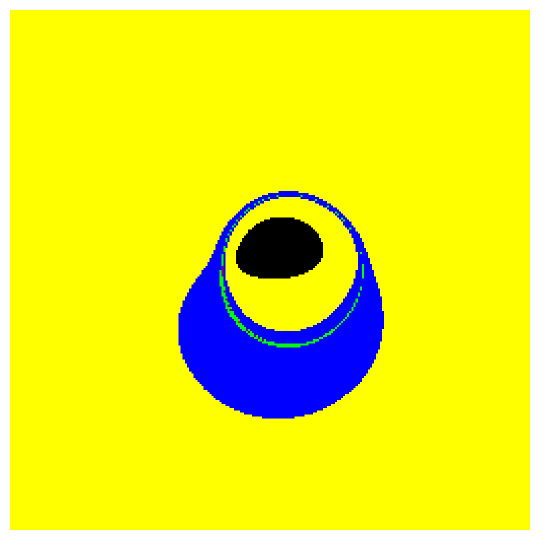}}
	\subfigure[$a=0.5,Q=0.1,\mathcal{G}=-0.1,\theta_o=75^\circ$]{\includegraphics[scale=0.4]{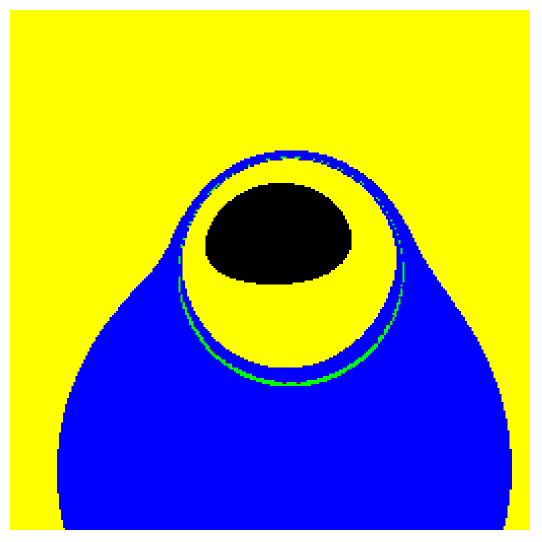}\label{fig6_12}}

	\caption{\label{fig6}The lensing bands for black holes under a prograde thin accretion disk. The black, yellow, blue, and green regions represent the inner shadow, direct image, lensed image, and higher-order image, respectively. The first to third columns correspond to observer inclinations $\theta_o = 0^\circ$, $17^\circ$, and $75^\circ$.}

\end{figure}

To better distinguish between the direct and lensed images, Figure~\ref{fig6} presents the lensing bands for different parameter settings. In the images, yellow, blue, and green denote the direct images, the lensed images, and the higher-order images. These colors correspond to the number of times photons cross the equatorial plane: yellow indicates one crossing, blue indicates two, and green indicates three. The black region at the center represents the inner shadow. Notably, the higher-order images always lie within the region of the lensed images. Inside the higher-order images, the intensity of the direct image is significantly higher than that of the lensed image. For the case $\theta_o = 0^\circ$ (the first column), the inner shadow appears as a perfect circle, while the direct and lensed images form concentric rings centered at the same point. The variation of $a$ and $Q$ affects the size of the inner shadow and the extent of the lensed image but does not alter their geometric shape. In particular, a decrease in $\mathcal{G}$ significantly enlarges both the inner shadow and the lensed image. For $\theta_o = 17^\circ$ (the second column), the lensed image and higher-order images shift downward on the screen. When $\theta_o$ increases to $75^\circ$ (the third column), the inner shadow and the lensed image undergo significant deformation. The lensed image shifts further to the lower left of the screen, and its upper region contracts into a narrow band. Meanwhile, for the higher-order images, their lower part is more discernible than the upper part. These observations indicate that each of the parameters $(a, Q, \mathcal{G}, \theta_o)$ influences the distributions of both the direct and the lensed images, with the Lorentz-violating parameter $\mathcal{G}$ having the most pronounced effect.

\begin{figure}[!h]
	\centering 
	\subfigure[$Q=0.1,\mathcal{G}=0.15,a=0.1$]{\includegraphics[scale=0.5]{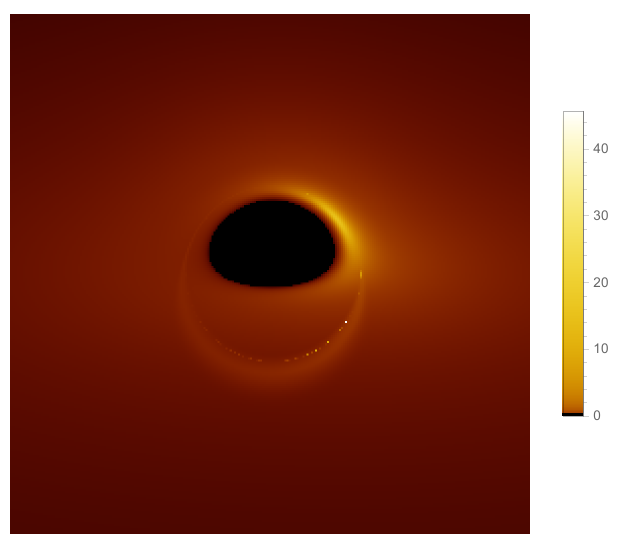}}
	\subfigure[$Q=0.1,\mathcal{G}=0.15,a=0.5$]{\includegraphics[scale=0.5]{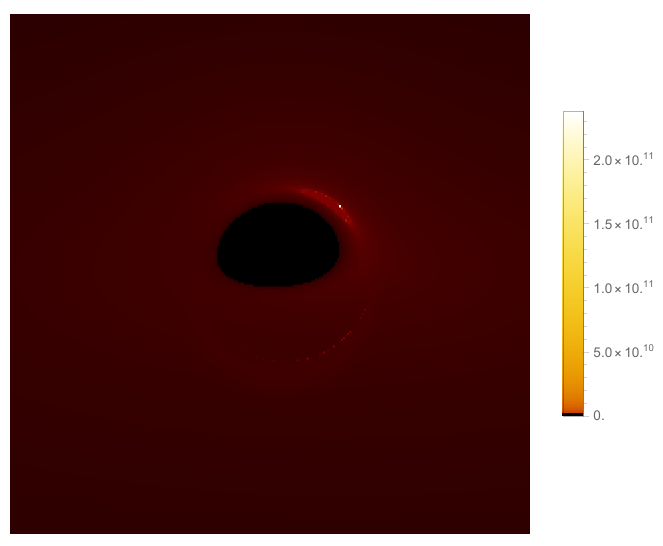}}
	\subfigure[$Q=0.1,\mathcal{G}=0.15,a=0.9$]{\includegraphics[scale=0.5]{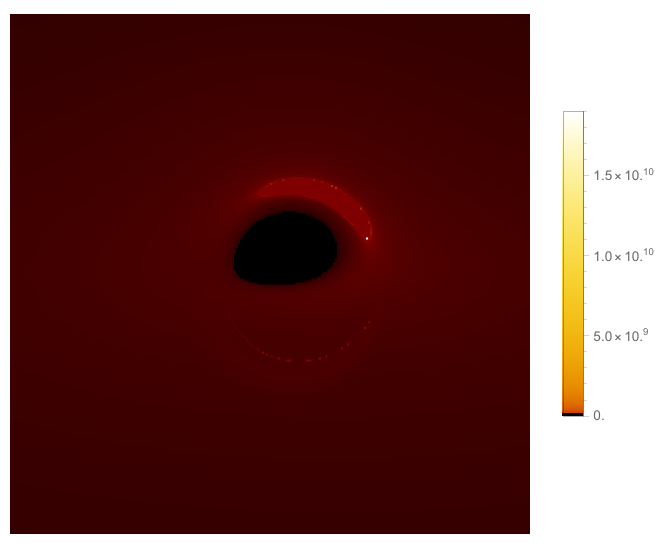}}
	\subfigure[$a=0.5,\mathcal{G}=0.15,Q=0.1$]{\includegraphics[scale=0.5]{Re2_1.pdf}}
	\subfigure[$a=0.5,\mathcal{G}=0.15,Q=0.3$]{\includegraphics[scale=0.5]{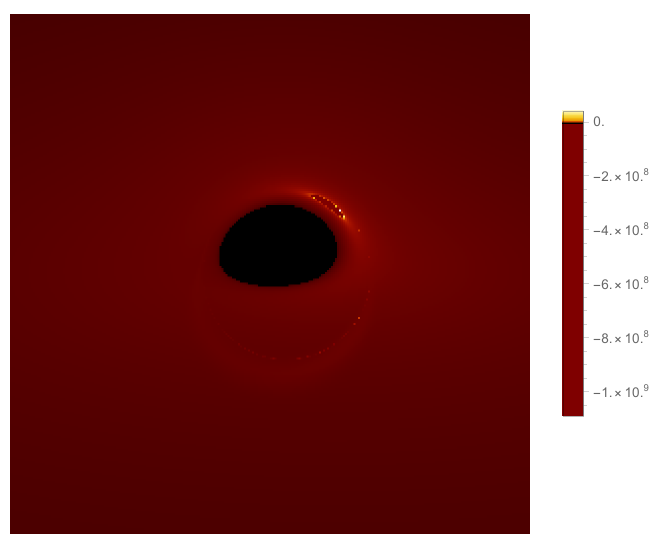}}
	\subfigure[$a=0.5,\mathcal{G}=0.15,Q=0.65$]{\includegraphics[scale=0.5]{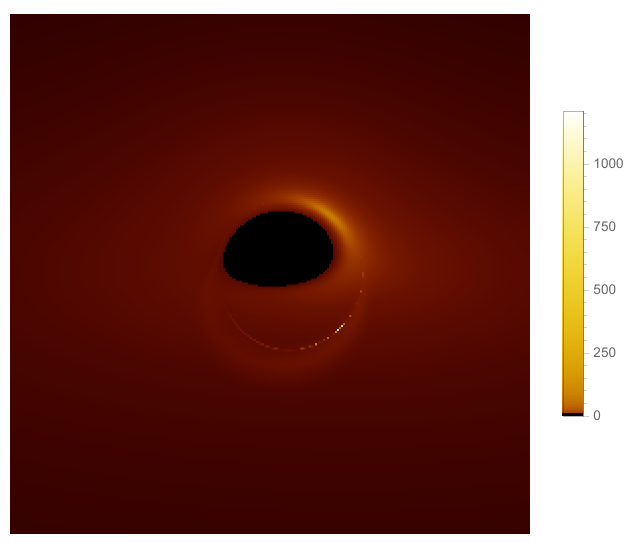}}
	\subfigure[$Q=0.1,a=0.5,\mathcal{G}=-0.1$]{\includegraphics[scale=0.5]{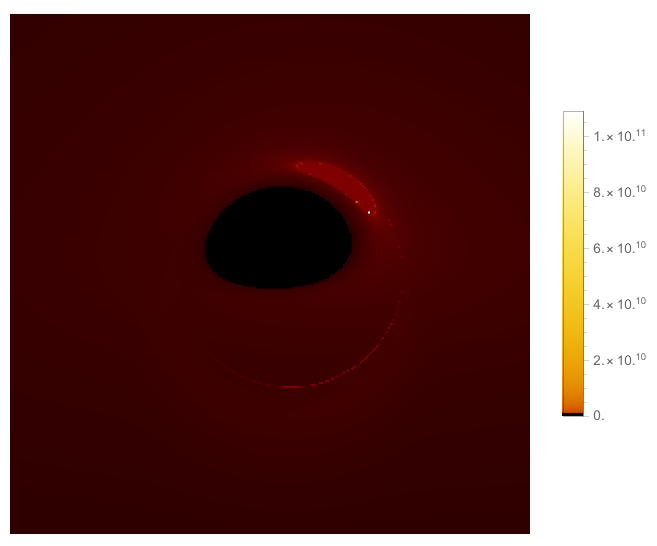}}
	\subfigure[$Q=0.1,a=0.5,\mathcal{G}=0.1$]{\includegraphics[scale=0.5]{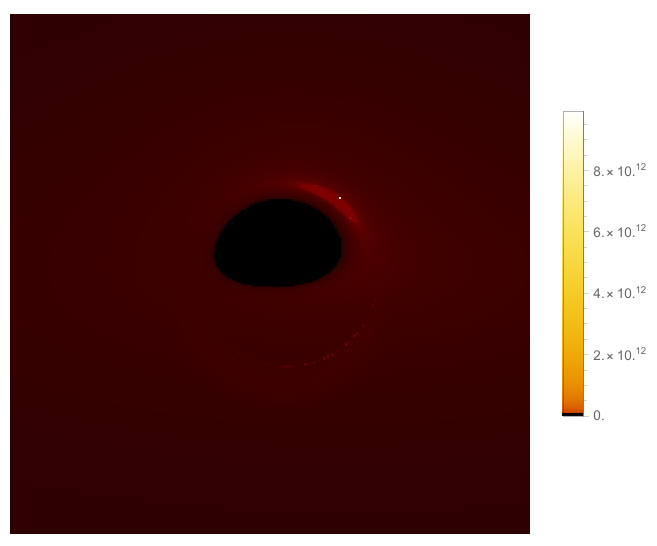}}
	\subfigure[$Q=0.1,a=0.5,\mathcal{G}=0.15$]{\includegraphics[scale=0.5]{Re2_1.pdf}}
	
	\caption{\label{fig7}Shadow images of black holes under a retrograde thin accretion disk. Top panel: rotation parameter $a$. Middle panel: charge $Q$. Bottom panel: Lorentz-violating parameter $\mathcal{G}$. The observer inclination is fixed at $\theta_o = 75^\circ$.}
\end{figure}

In the final part of this section, we turn to the case of retrograde accretion disks. Figure \ref{fig7} shows the influence of $(a, Q, \mathcal{G})$ on the shadow images of rotating charged black holes in KR gravity surrounded by retrograde accretion disks, with the observer's inclination angle fixed at $\theta_o = 75^\circ$. As shown in the figure, the inner shadow persists across the entire parameter space, and the effect of the parameters on its size is similar to that in the prograde case. However, in contrast to the prograde scenario, gravitational redshift significantly reduces the observed brightness of the shadow images. The overall decrease in intensity makes it more difficult to distinguish between the lensed and higher-order images, and also diminishes the visibility of the critical curve. Notably, for retrograde accretion disks, the crescent-shaped bright area appears on the left side of the image plane. This shift results from a change in the motion direction of the accreting material, which alters the direction of the blueshift caused by the Doppler effect. In addition, an increase in $Q$ enhances the overall brightness of the image, while a decrease in $\mathcal{G}$ enlarges the extent of the crescent-shaped bright region.

\section{Distribution of Redshift Factors}\label{sec4}

The distribution of intensity on the imaging plane is affected by several physical mechanisms, including photon divergence, absorption, and redshift. In this section, we focus on the distribution of the redshift factor $\chi_n$ on the imaging plane. Redshift is known to arise solely from two mechanisms: gravitational redshift and Doppler redshift. The former is always present, while the latter originates from the relative motion between the accretion disk and the observer. A comparison of the two mechanisms shows that the Doppler effect plays a dominant role in explaining the intensity distribution in the image, whereas gravitational redshift reflects the influence of strong gravitational fields on photon propagation and also has a noticeable impact. Therefore, accurately computing the redshift factor induced by moving particles is crucial in the imaging process of black hole shadows, as it plays a key role in characterizing the physical properties of the black hole and its accretion disk.

\begin{figure}[!h]
	\centering 
	\subfigure[$a=0.5,Q=0.1,\mathcal{G}=0.15,\theta_o=0^\circ$]{\includegraphics[scale=0.35]{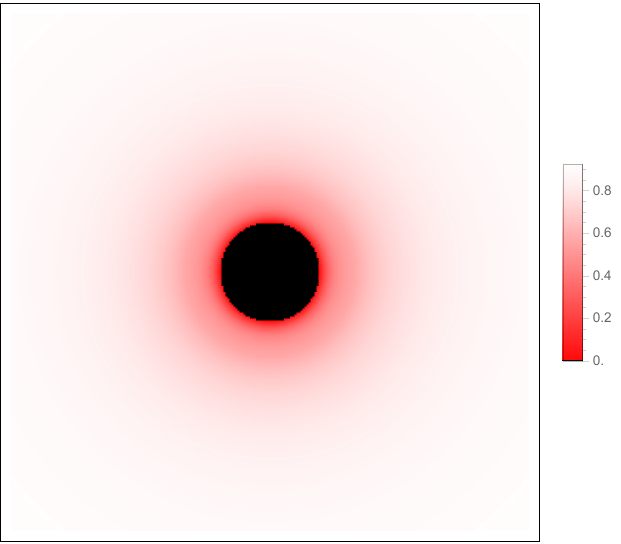}}
	\subfigure[$a=0.9,Q=0.1,\mathcal{G}=0.15,\theta_o=0^\circ$]{\includegraphics[scale=0.35]{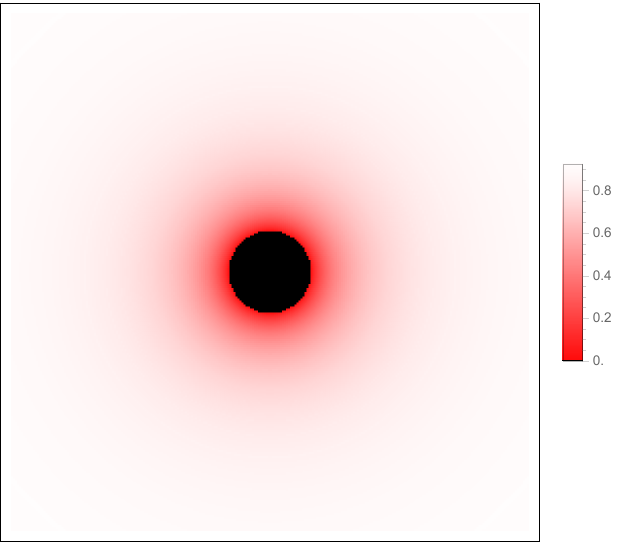}}
	\subfigure[$a=0.5,Q=0.65,\mathcal{G}=0.15,\theta_o=0^\circ$]{\includegraphics[scale=0.35]{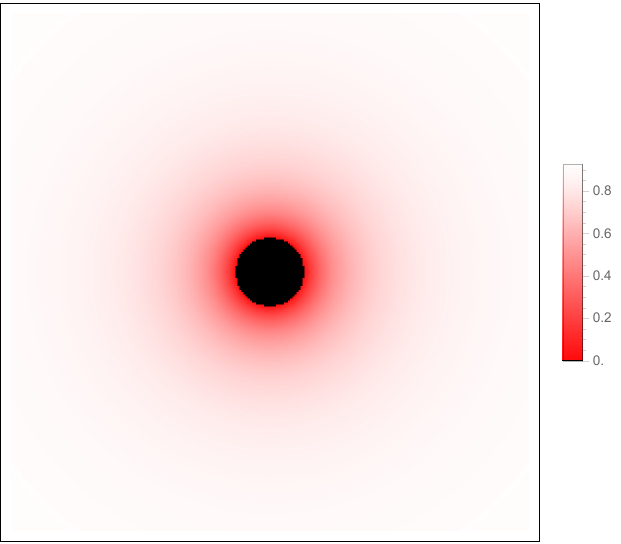}}
	\subfigure[$a=0.5,Q=0.1,\mathcal{G}=-0.1,\theta_o=0^\circ$]{\includegraphics[scale=0.35]{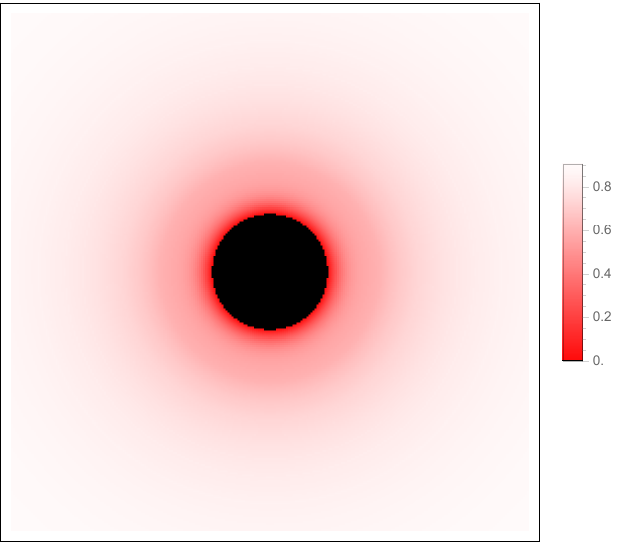}}
	\subfigure[$a=0.5,Q=0.1,\mathcal{G}=0.15,\theta_o=17^\circ$]{\includegraphics[scale=0.35]{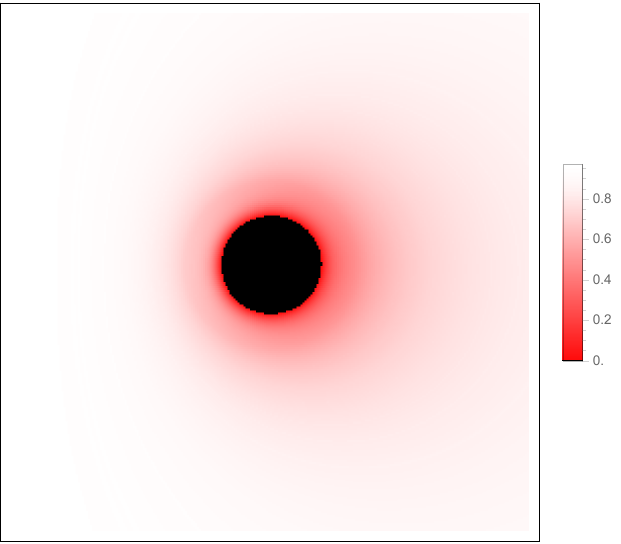}}
	\subfigure[$a=0.9,Q=0.1,\mathcal{G}=0.15,\theta_o=17^\circ$]{\includegraphics[scale=0.35]{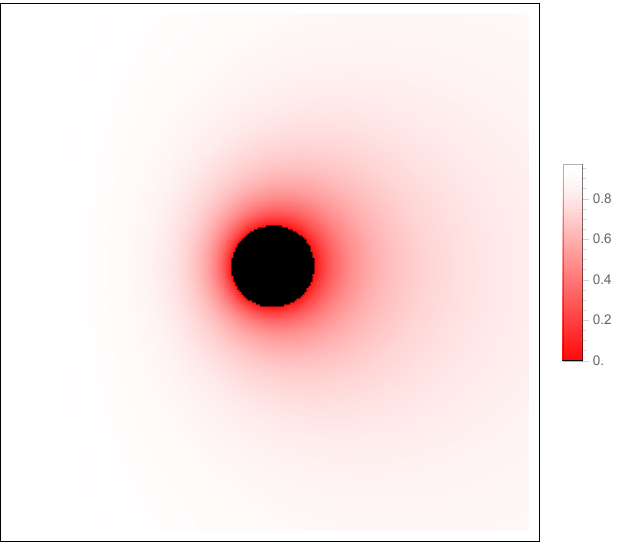}}
	\subfigure[$a=0.5,Q=0.65,\mathcal{G}=0.15,\theta_o=17^\circ$]{\includegraphics[scale=0.35]{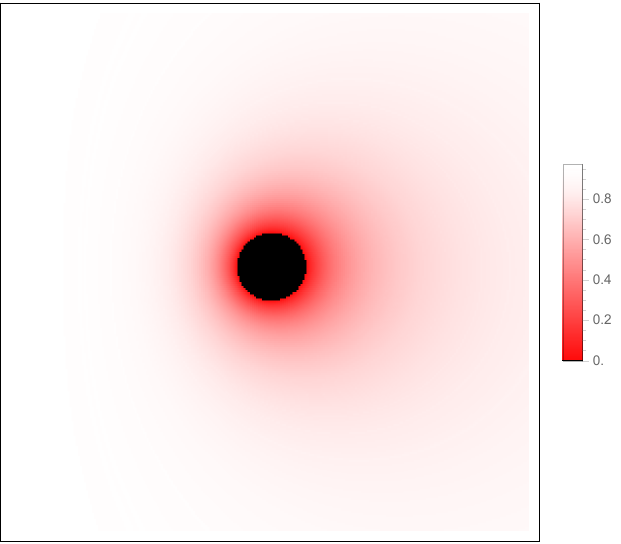}}
	\subfigure[$a=0.5,Q=0.1,\mathcal{G}=-0.1,\theta_o=17^\circ$]{\includegraphics[scale=0.35]{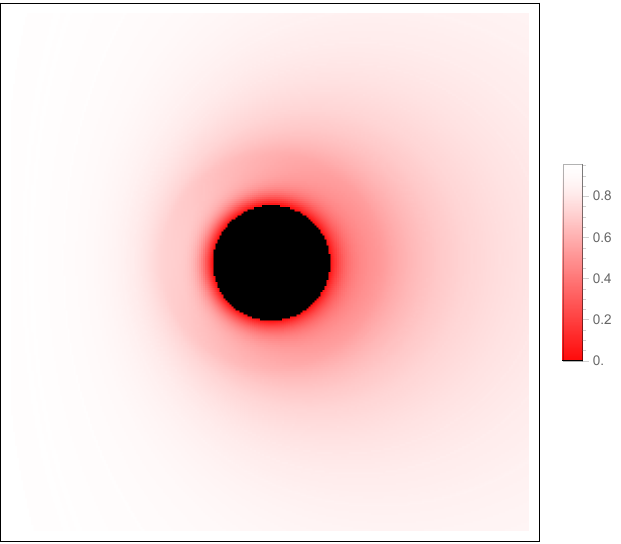}}
	\subfigure[$a=0.5,Q=0.1,\mathcal{G}=0.15,\theta_o=75^\circ$]{\includegraphics[scale=0.35]{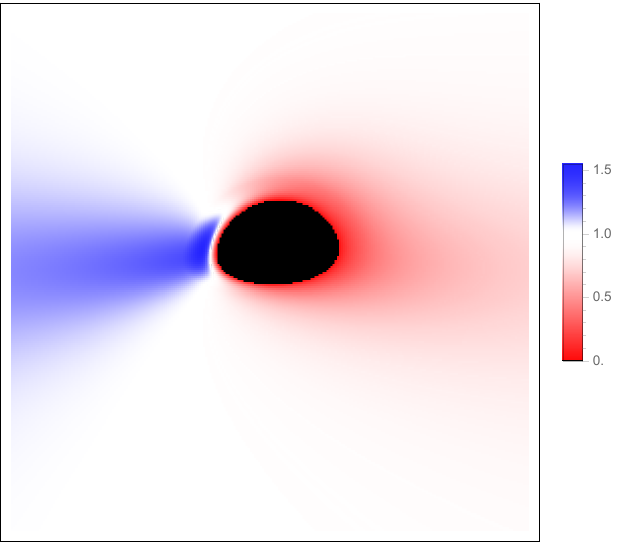}}
	\subfigure[$a=0.9,Q=0.1,\mathcal{G}=0.15,\theta_o=75^\circ$]{\includegraphics[scale=0.35]{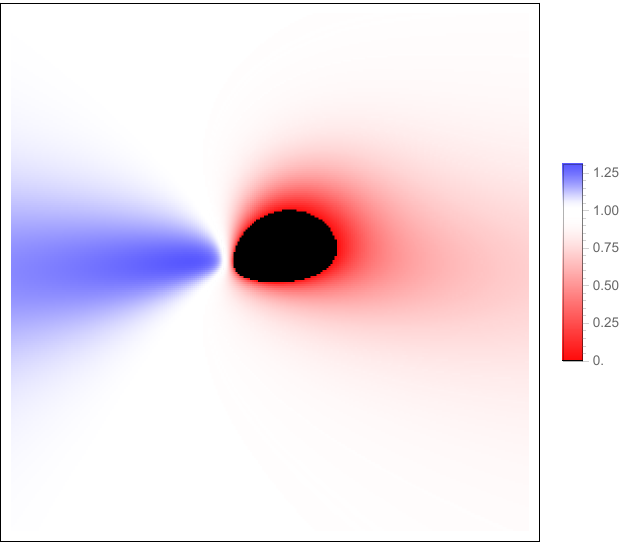}}
	\subfigure[$a=0.5,Q=0.65,\mathcal{G}=0.15,\theta_o=75^\circ$]{\includegraphics[scale=0.35]{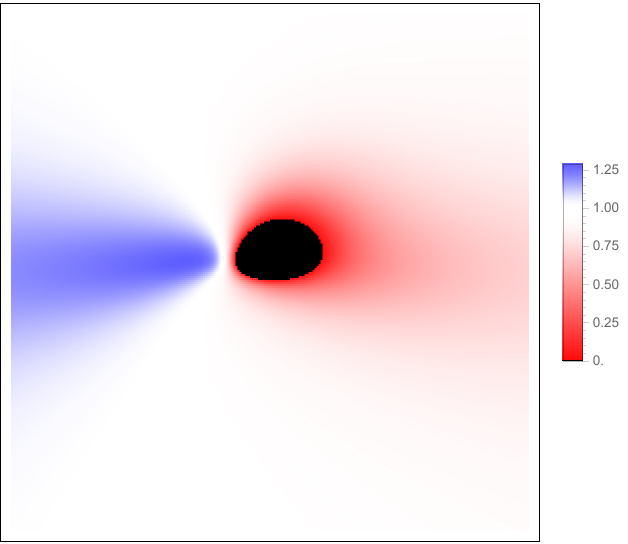}}
	\subfigure[$a=0.5,Q=0.1,\mathcal{G}=-0.1,\theta_o=75^\circ$]{\includegraphics[scale=0.35]{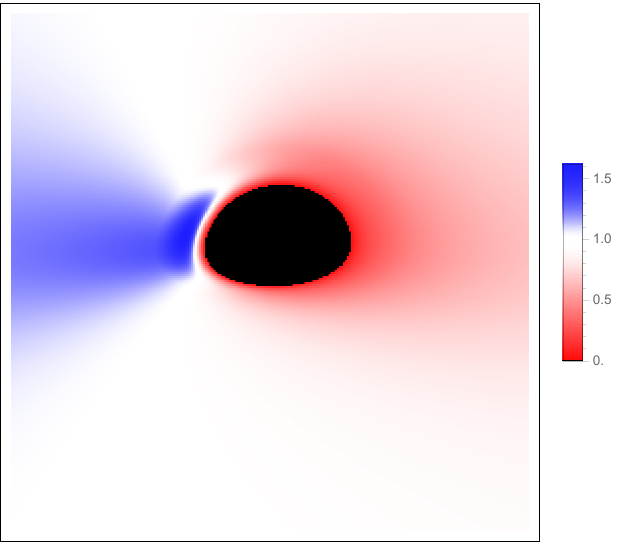}}
	
	\caption{\label{fig8}Redshift factor distributions for the direct images under a prograde thin accretion disk. The black region represents the inner shadow while red and blue denote redshift and blueshift respectively, with deeper colors linearly indicating stronger effects. The first to third columns correspond to observer inclinations $\theta_o = 0^\circ$, $17^\circ$, and $75^\circ$.}
\end{figure}

\begin{figure}[!h]
	\centering 
	\subfigure[$a=0.5,Q=0.1,\mathcal{G}=0.15,\theta_o=0^\circ$]{\includegraphics[scale=0.35]{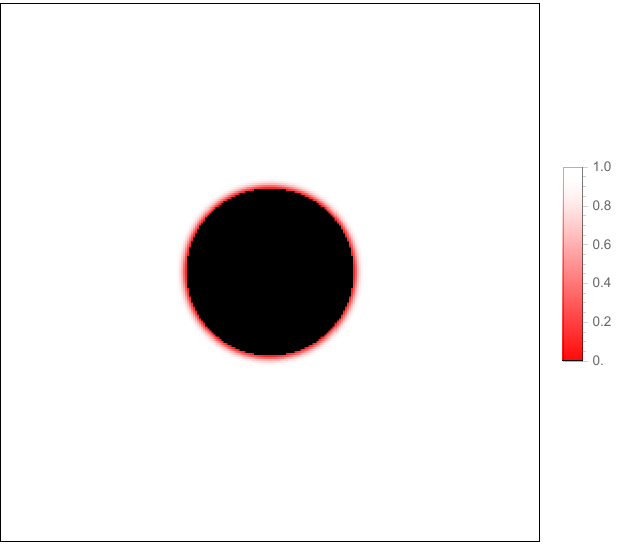}}
	\subfigure[$a=0.9,Q=0.1,\mathcal{G}=0.15,\theta_o=0^\circ$]{\includegraphics[scale=0.35]{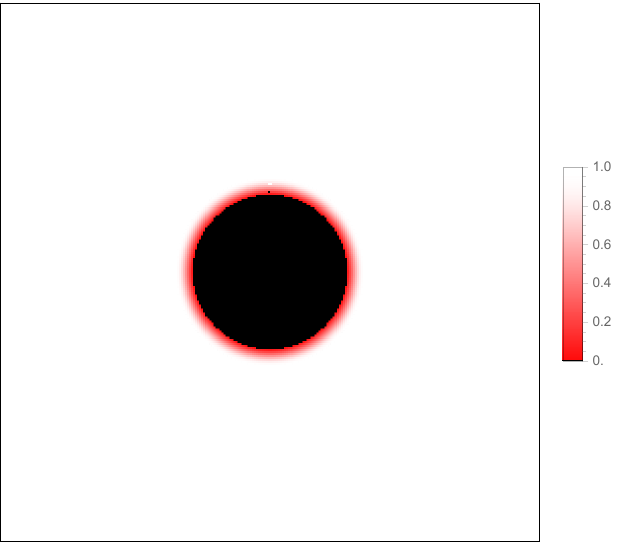}}
	\subfigure[$a=0.5,Q=0.65,\mathcal{G}=0.15,\theta_o=0^\circ$]{\includegraphics[scale=0.35]{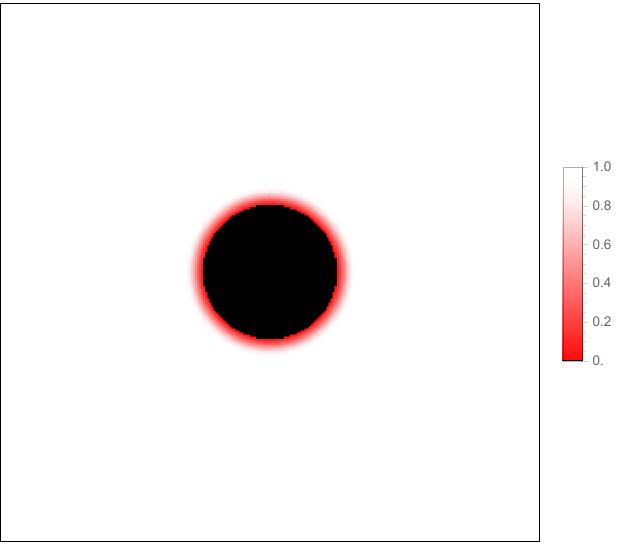}}
	\subfigure[$a=0.5,Q=0.1,\mathcal{G}=-0.1,\theta_o=0^\circ$]{\includegraphics[scale=0.35]{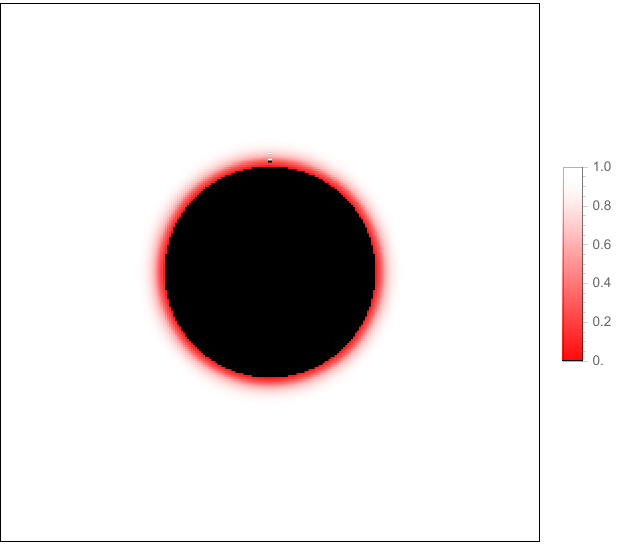}}
	\subfigure[$a=0.5,Q=0.1,\mathcal{G}=0.15,\theta_o=17^\circ$]{\includegraphics[scale=0.35]{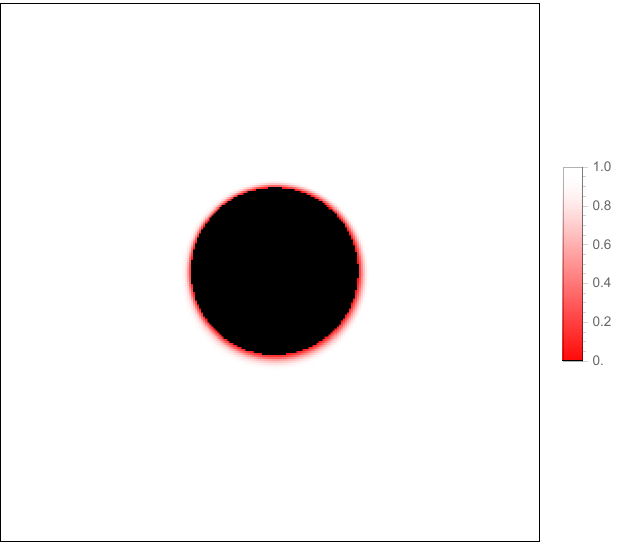}}
	\subfigure[$a=0.9,Q=0.1,\mathcal{G}=0.15,\theta_o=17^\circ$]{\includegraphics[scale=0.35]{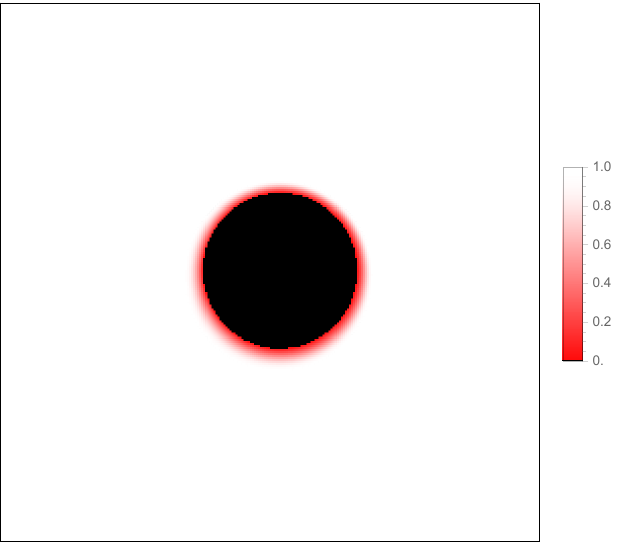}}
	\subfigure[$a=0.5,Q=0.65,\mathcal{G}=0.15,\theta_o=17^\circ$]{\includegraphics[scale=0.35]{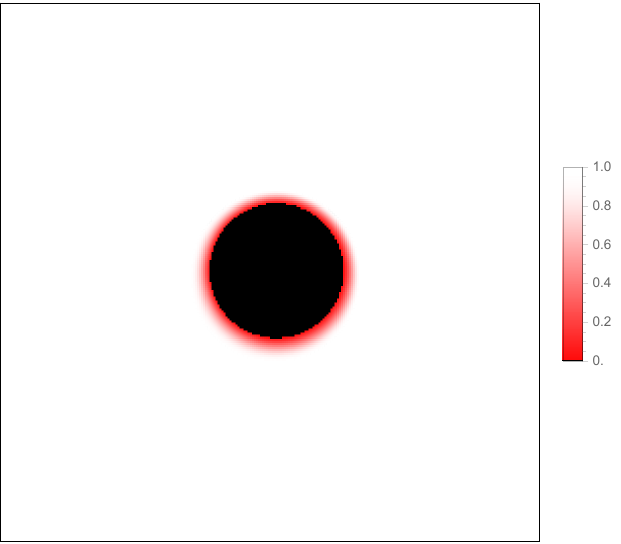}}
	\subfigure[$a=0.5,Q=0.1,\mathcal{G}=-0.1,\theta_o=17^\circ$]{\includegraphics[scale=0.35]{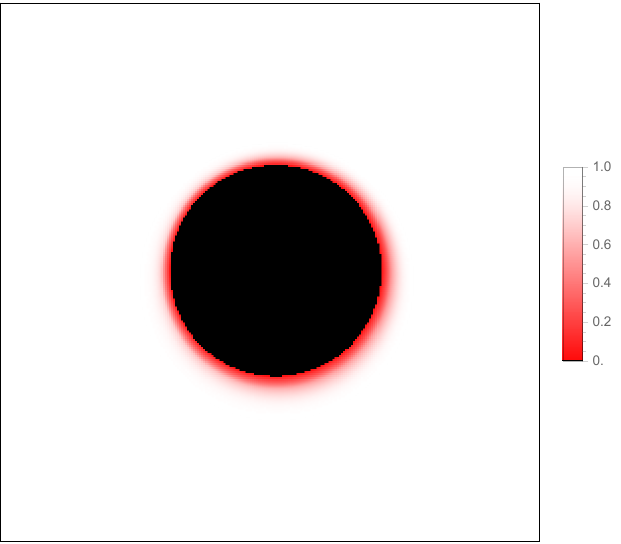}}
	\subfigure[$a=0.5,Q=0.1,\mathcal{G}=0.15,\theta_o=75^\circ$]{\includegraphics[scale=0.35]{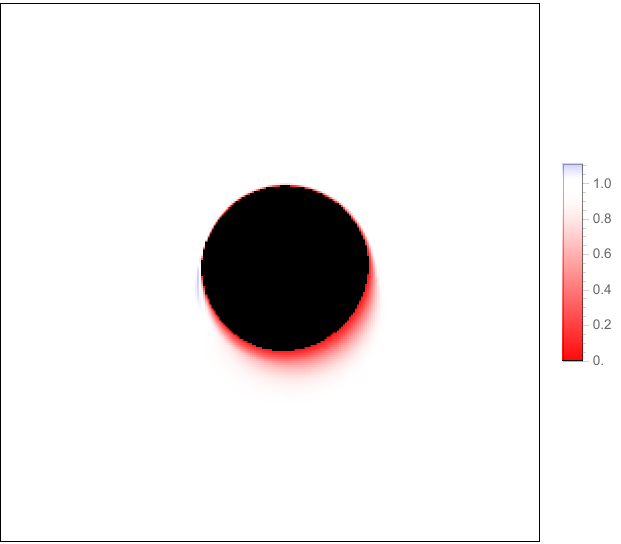}}
	\subfigure[$a=0.9,Q=0.1,\mathcal{G}=0.15,\theta_o=75^\circ$]{\includegraphics[scale=0.35]{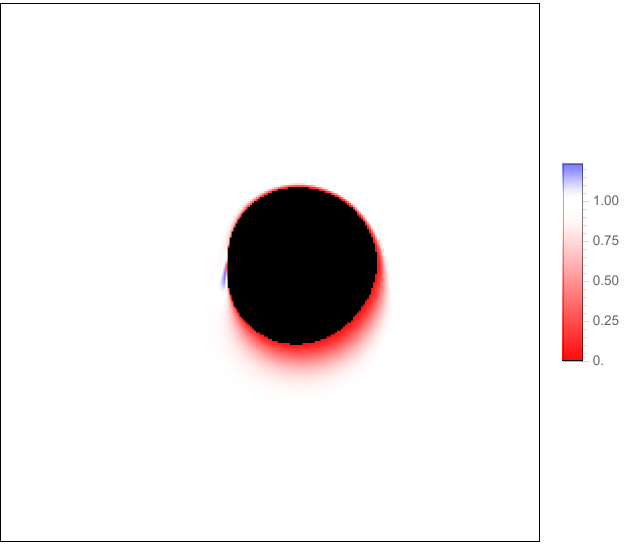}}
	\subfigure[$a=0.5,Q=0.65,\mathcal{G}=0.15,\theta_o=75^\circ$]{\includegraphics[scale=0.35]{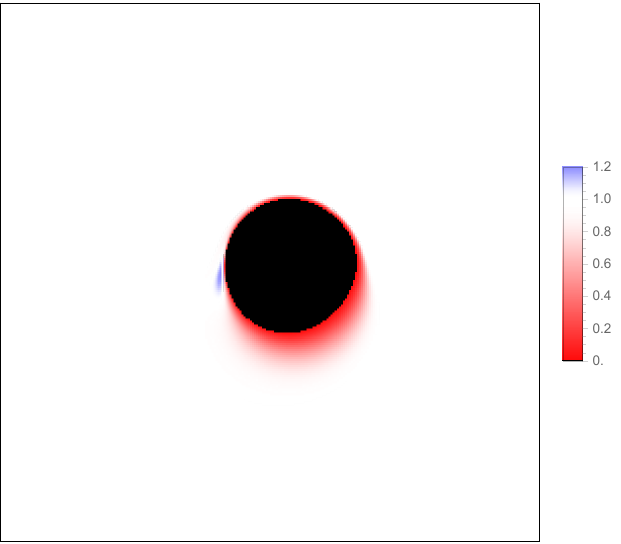}}
	\subfigure[$a=0.5,Q=0.1,\mathcal{G}=-0.1,\theta_o=75^\circ$]{\includegraphics[scale=0.35]{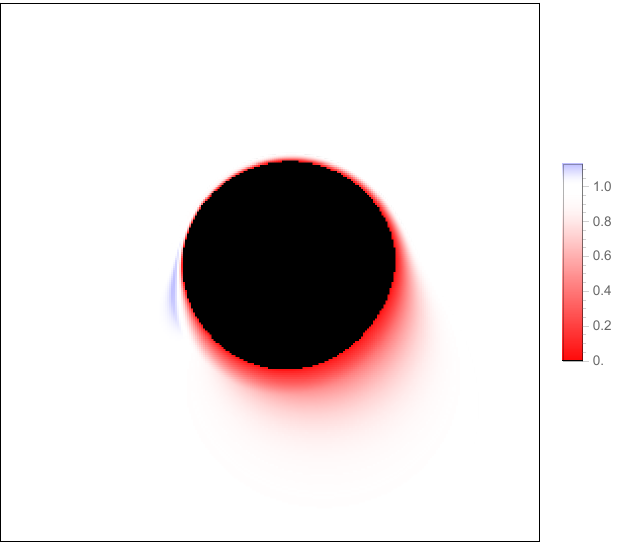}}
	
	\caption{\label{fig9}Redshift factor distributions for the lensed images under a prograde thin accretion disk. Red and blue denote redshift and blueshift respectively, with deeper colors linearly indicating stronger effects. The first to third columns correspond to observer inclinations $\theta_o = 0^\circ$, $17^\circ$, and $75^\circ$.}
\end{figure}

Figures \ref{fig8} and \ref{fig9} show the distribution of redshift factors for the direct and lensed images under prograde accretion disks, with parameter values consistent with those used in Figure \ref{fig6}. The redshift factors $\chi_1$ and $\chi_2$, corresponding to the direct and lensed images, are computed via (\ref{rsf1}) and (\ref{rsf2}). In the images, red and blue denote redshift and blueshift, with deeper colors linearly indicating stronger effects. 

For the direct image shown in Figure \ref{fig8}, it can be seen that when the observer inclination is $\theta_o = 0^\circ$ or $17^\circ$, only redshift appears, with no blueshift. The redshift becomes more pronounced near the inner shadow. In this case, variations in the parameters $(a, Q, \mathcal{G})$ have little effect on the overall distribution of the redshift factor. When $\theta_o = 75^\circ$, a clear blueshift appears on the left side of the image, and its strength increases progressively near the inner shadow, corresponding precisely to the crescent-shaped bright area in the intensity distribution. In this scenario, increasing $a$ or $Q$ causes the redshift and blueshift regions to separate, while decreasing $\mathcal{G}$ brings them closer together. At low inclination angles, the redshift is primarily due to gravitational effects. As the inclination increases, the relative radial velocity becomes larger, which enhances the Doppler effect and leads to a blueshift on the left side of the image. It is worth noting that the appearance of blueshift on the left and redshift on the right is determined by the direction of motion of the prograde accretion disk. If the disk were retrograde, these two regions would be reversed. For the lensed image shown in Figure \ref{fig9}, the redshift distribution outside the black region forms a concentrated red ring. When $\theta_o$ increases to $75^\circ$, the redshift region shifts toward the lower right, and a slight blueshift appears on the left side of the image, which is also caused by the Doppler effect.

\section{Polarization Images under Synchrotron Radiation}\label{sec5}

Synchrotron radiation is an important emission mechanism in the accretion disks surrounding black holes, generated by hot gas and characterized by electromagnetic radiation with a high degree of polarization. This property makes it particularly valuable for studying the polarization images of black holes. In this context, this section explores the polarization images of rotating charged black holes in KR gravity under synchrotron radiation. To enhance image visualization, we incorporate the polarization images with the thin accretion disk model.

For an observer comoving with the plasma, whose four-velocity is $u^\mu$, the polarization direction $\vec{f}$ of the emitted light is always perpendicular to both the local magnetic field $\vec{B}$ and the photon's wave vector $\vec{k}$,
\begin{equation}
	\vec{f} = \frac{\vec{k} \times \vec{B}}{|\vec{k}||\vec{B}|}.
\end{equation}
The generally covariant form of the above equation is
\begin{equation}
	f^\mu \propto \epsilon^{\mu\nu\alpha\beta} u_\nu k_\alpha \vec{B}_\beta.
\end{equation}
In the imaging process, it is necessary to determine the direction of the polarization vector and normalize it so that it satisfies the orthonormality condition,
\begin{equation}
	f^\mu f_\mu = 1.
\end{equation}
The intensities of linearly polarized and unpolarized light at the emission point are denoted by the emission functions $\mathcal{E}_p$ and $\mathcal{E}_i$, respectively. For simplicity, we assume that the emission intensities are independent of photon frequency and magnetic field, and depend only on position, so that
\begin{equation}
	\mathcal{E}_i = \mathcal{E}_i(r), \qquad \mathcal{E}_p = \xi \mathcal{E}_i(r),
\end{equation}
where $\xi \in [0, 1]$ characterizes the fraction of linearly polarized light at the emission point. If the emitted light is assumed to be fully linearly polarized, then $\xi = 1$. In the geometric optics approximation, the polarization vector $f^\mu$ undergoes parallel transport along the photon geodesic,
\begin{equation}
	k^\nu \nabla_\nu f^\mu = 0.
\end{equation}
This can be rewritten as
\begin{equation}
	\frac{d}{d\lambda} f^\mu + \Gamma^{\mu}{}_{\nu\alpha} k^\nu f^\alpha = 0,
\end{equation}
where $\lambda$ is the affine parameter. At the observer, the linearly polarized intensity $\mathcal{P}_{\nu_{o}}$ and the total intensity $I_o$ are given by the same expressions as in the unpolarized case,
\begin{equation}
	\mathcal{P}_{\nu_{o}} = \chi^3 \mathcal{E}_p, \qquad I_o = \chi^3 \mathcal{E}_i,
\end{equation}
where $\chi$ is the redshift factor. In (\ref{eq:zamo}), we construct a set of orthogonal tetrads (ZAMO) at the observer's position, which are used to define the imaging screen. On the screen, we select a set of orthonormal vectors $(e_{(\theta)}, e_{(\varphi)})$ as the basis, and the projection of the polarization vector onto the imaging plane can be expressed as
\begin{equation}
	f^{(\alpha)} = f^\mu \cdot e_\alpha = -f^\mu \cdot e_\varphi, \qquad f^{(\beta)} = f^\mu \cdot e_\beta = -f^\mu \cdot e_\theta.
\end{equation}
After determining the direction of the polarization vector, the total linearly polarized intensity can be obtained by summing the linearly polarized intensities over all emission points on the equatorial plane. According to the definitions of the Stokes parameters $\mathcal{Q}$ and $\mathcal{U}$~\cite{huang2024coport}, which obey the principle of linear superposition, the final results can be obtained by summing $\mathcal{Q}$ and $\mathcal{U}$,
\begin{equation}
	\mathcal{Q}_{all} = \sum_{n=1}^{N} \chi_n^3 \mathcal{E}_{pn} \left[ \left( f_n^{(\alpha)} \right)^2 - \left( f_n^{(\beta)} \right)^2 \right], \quad
	\mathcal{U}_{all} = \sum_{n=1}^{N} \chi_n^3 \mathcal{E}_{pn} \left( 2 f_n^{(\alpha)} f_n^{(\beta)} \right),
\end{equation}
where $n = 1, \ldots, N$ denotes the number of intersections of the light ray with the equatorial plane. The total linearly polarized intensity and the electric vector position angle (EVPA) at the observer are
\begin{equation}
	\mathcal{P}_o = \sqrt{ \mathcal{Q}_{all}^2 + \mathcal{U}_{all}^2 }, \quad \Psi = \frac{1}{2} \arctan \frac{ \mathcal{U}_{all} }{ \mathcal{Q}_{all} }.
\end{equation}
Here, we adopt the gauge $f^{(\beta)} > 0, \Psi \in (0, \pi)$.

\begin{figure}[!h]
	\centering 

	\subfigure[$Q=0.1,\mathcal{G}=0.15,a=0.1$]{\includegraphics[scale=0.5]{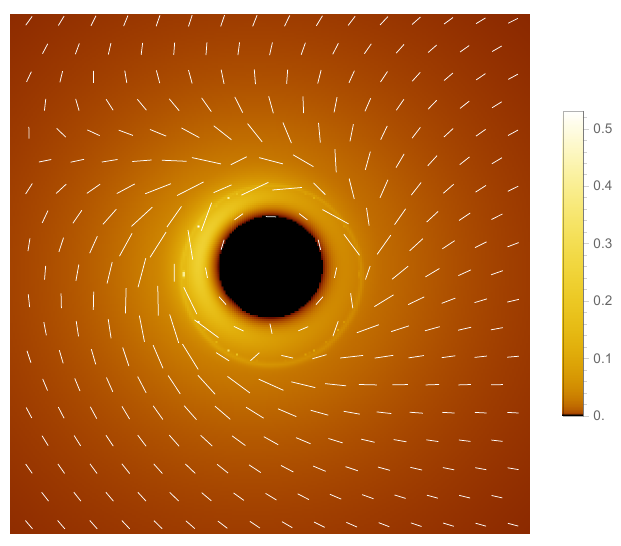}}
	\subfigure[$Q=0.1,\mathcal{G}=0.15,a=0.5$]{\includegraphics[scale=0.5]{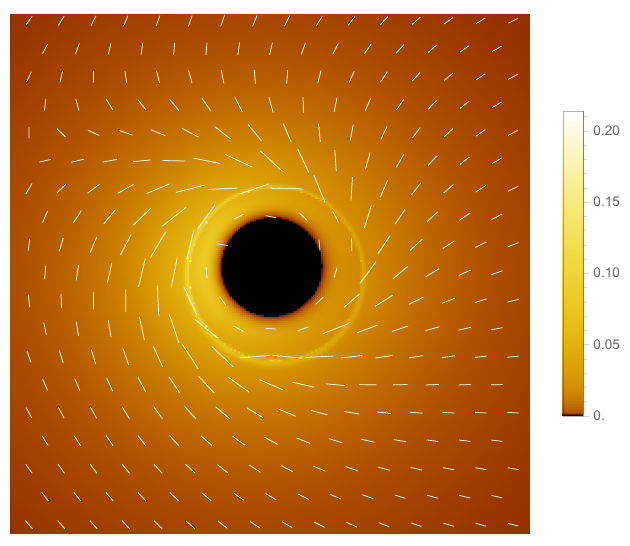}}
	\subfigure[$Q=0.1,\mathcal{G}=0.15,a=0.9$]{\includegraphics[scale=0.5]{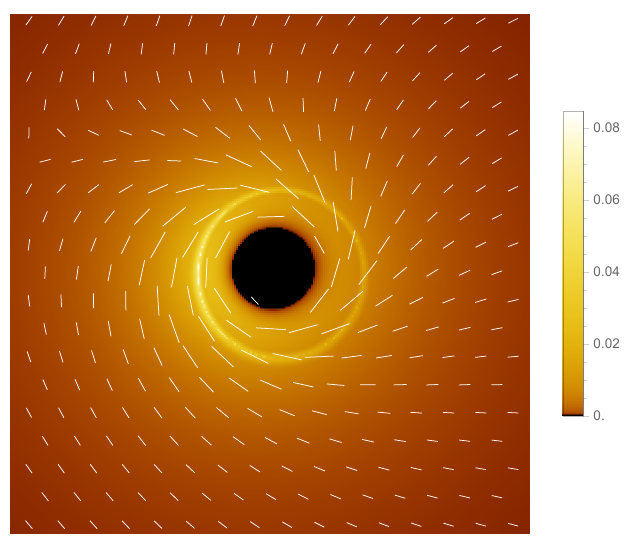}}

	\subfigure[$a=0.5,\mathcal{G}=0.15,Q=0.1$]{\includegraphics[scale=0.5]{Po2.pdf}}
	\subfigure[$a=0.5,\mathcal{G}=0.15,Q=0.2$]{\includegraphics[scale=0.5]{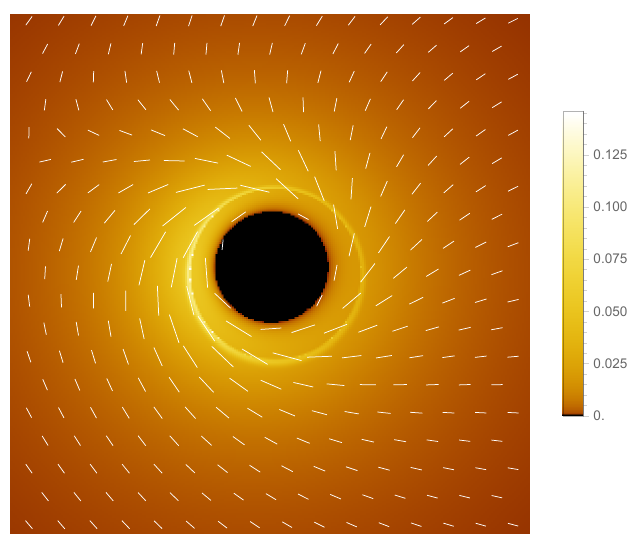}}
	\subfigure[$a=0.5,\mathcal{G}=0.15,Q=0.3$]{\includegraphics[scale=0.5]{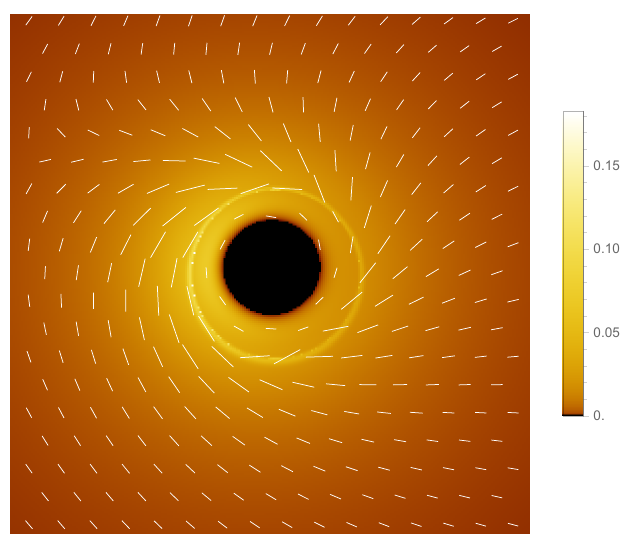}}

	\subfigure[$Q=0.1,a=0.5,\mathcal{G}=-0.1$]{\includegraphics[scale=0.5]{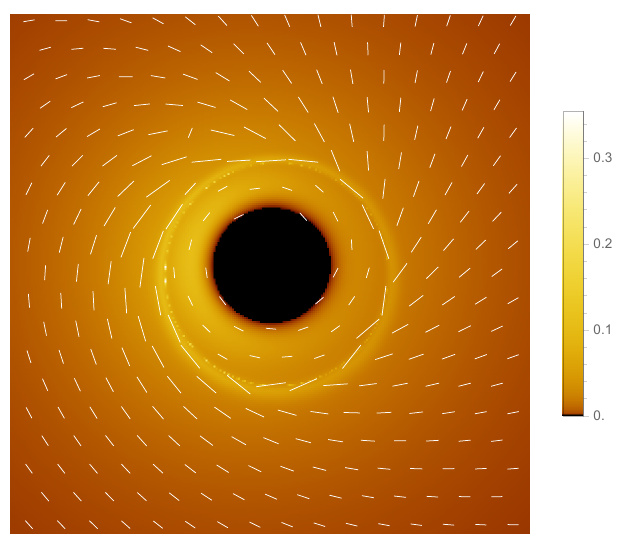}}
	\subfigure[$Q=0.1,a=0.5,\mathcal{G}=0$\label{fig10_8}]{\includegraphics[scale=0.5]{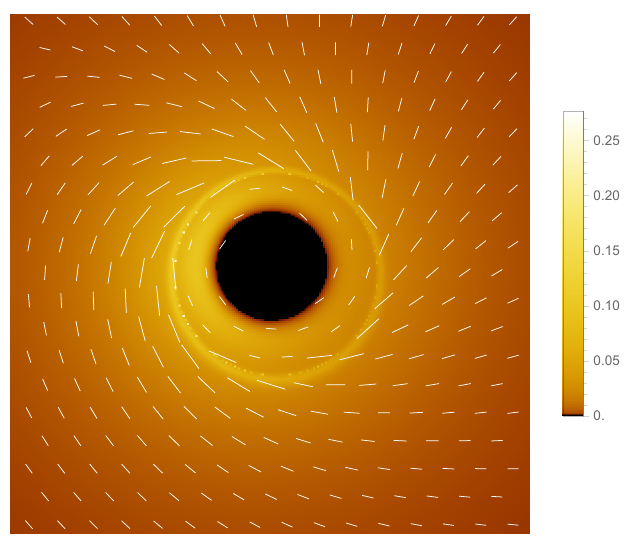}}
	\subfigure[$Q=0.1,a=0.5,\mathcal{G}=0.15$]{\includegraphics[scale=0.5]{Po2.pdf}}
	\caption{\label{fig10}Polarization images of black holes generated by synchrotron radiation. White line segments indicate linear polarization vectors with varying intensities. Top panel: rotation parameter $a$. Middle panel: charge $Q$. Bottom panel: Lorentz-violating parameter $\mathcal{G}$. All other parameters are fixed with observer inclination $\theta_o = 17^\circ$ and magnetic field $\vec{B} = (0.87, 0.5, 0)$.}
\end{figure}

Figure~\ref{fig10} presents the numerical simulation results of the polarization images. For M87*, the observer inclination is taken as $\theta_o = 17^\circ$. Assuming a Schwarzschild black hole, the optimal magnetic field orientation can be estimated as $\vec{B} = (0.87, 0.5, 0)$. Accordingly, the same values of $\theta_o$ and $\vec{B}$ are adopted in our simulations. The background image corresponds to the thin accretion disk model, whose details have been described earlier. White line segments indicate the linear polarization vectors, with their length representing the polarization intensity $\mathcal{P}_o$. The direction of each line segment corresponds to the electric vector position angle $\Psi$.

In Figure~\ref{fig10}, the first row reflects the variation of $a$, the second row reflects the variation of $Q$, and the third row reflects the variation of $\mathcal{G}$ (where Figure~\ref{fig10_8} corresponds to the Kerr–Newman black hole case). In the images, the polarization intensity $\mathcal{P}_o$ reaches its maximum near the lensed image and higher-order images. This is because photons that intersect the equatorial plane more frequently correspond to larger values of $N$, which leads to higher values of $\mathcal{P}_o$. Beyond the regions of the lensed images, $\mathcal{P}_o$ rapidly drops to zero. Due to the presence of the event horizon, no polarization vectors exist within the inner shadow. This stands in sharp contrast to other compact objects without event horizons—for instance, in the case of boson stars, polarization vectors can appear within the interior region~\cite{li2025observational}. Such features may serve as potential observational signatures for distinguishing black holes from horizonless compact objects. 

\section{Conclusion and Discussion}\label{sec6}

The breakthrough in imaging technology achieved by the EHT marks the beginning of a new era in testing gravity in the strong-field regime. The formation of black hole shadows fundamentally arises from the deflection of light in strong gravitational fields, and their observable features help deepen our understanding of the underlying spacetime geometry. In recent years, the study of various types of black hole shadows has progressed rapidly, offering new possibilities for testing a wide range of gravitational theories. In this paper, we systematically investigated the shadow images of rotating charged black holes in KR gravity under the thin accretion disk model, with a particular focus on the effects of the observer inclination $\theta_o$, the rotation parameter $a$, the charge $Q$, and the Lorentz-violating parameter $\mathcal{G}$.

To quantify the observational characteristics of rotating charged black holes in KR gravity, we first calculated the angular diameter of the shadow and constrained the parameter space by comparing the results with EHT observational data. The results indicate that, compared to Sgr A*, the observational constraints from M87* impose stronger bounds on the parameters $Q$ and $\mathcal{G}$ in KR gravity. When $Q$ is large or $\mathcal{G}$ is small, the estimated angular diameter of the M87* shadow exceeds the $1\sigma$ confidence interval. For Sgr A*, although variations in the parameters affect the range of angular diameter estimates, the values remain within the $1\sigma$ observational confidence interval.

We then focused on the shadow images of black holes illuminated by an accretion disk located on the equatorial plane, which is assumed to be both optically and geometrically thin, and presented both the intensity distributions and the lensing bands. The analysis concentrated on the features of the inner shadow, the direct image, the lensed image, and the critical curve. At low observer inclinations $\theta_o$, the direct and lensed images are difficult to distinguish. At higher inclinations, a crescent-shaped bright area appears on the left side of the inner shadow due to the Doppler effect, and the inner shadow also becomes deformed as $\theta_o$ increases. An increase in the rotation parameter $a$ reduces the size of the inner shadow and enhances the brightness of the critical curve. A larger charge $Q$ slightly decreases the size of the critical curve and improves its visibility. Interestingly, when the Lorentz-violating parameter $\mathcal{G}$ is positive, the bright band outside the inner shadow becomes more concentrated and smaller than when $\mathcal{G}$ is negative. Regarding the lensing bands, it is noteworthy that at a high observer inclination ($\theta_o = 75^\circ$), decreasing $\mathcal{G}$ significantly enlarges the radiation range of the lensed image. We further studied the shadow images under retrograde accretion disks. Due to gravitational redshift, the overall brightness decreases, making it more difficult to distinguish the lensed and higher-order images. Unlike the prograde case, the crescent-shaped bright area appears on the right side of the image and its size is negatively correlated with $\mathcal{G}$.

Building on this, we plotted the redshift factor distributions for both the direct and lensed images. At smaller observer inclinations ($\theta_o = 0^\circ$ and $17^\circ$), only redshift appears in both cases and the effect is primarily governed by gravitational redshift. The redshift becomes increasingly concentrated near the inner shadow, originating from the light emitted by particles plunging into the black hole along critical trajectories. As the observer inclination increases, the Doppler effect becomes stronger, leading to a prominent blueshift on the right side of the direct image, corresponding to the crescent-shaped bright area in the intensity distribution. Under such conditions, variations in the parameters $(a, Q, \mathcal{G})$ affect the boundary between redshifted and blueshifted regions. A larger $\theta_o$ also induces a slight blueshift on the left side of the lensed image. Unlike the direct image, the redshift in the lensed image is entirely concentrated near the edge of the black region.

Finally, we examined the polarization images under synchrotron radiation by introducing two observable physical quantities: the polarization intensity $\mathcal{P}_o$ and the electric vector position angle $\Psi$. The results show that $\mathcal{P}_o$ reaches its maximum near the higher-order images and rapidly drops to zero beyond the regions of the direct and lensed images. In contrast, $\Psi$ shows weak dependence on $(a, \mathcal{G}, Q)$ but is more significantly influenced by its relative position on the imaging plane.

Our study highlights the influence of black hole parameters, particularly the Lorentz-violating parameter $\mathcal{G}$, on shadow features under the thin accretion disk model. While placing constraints on the parameter space, it also provides a theoretical basis for distinguishing rotating charged black holes in KR gravity from other types of black holes through future high-resolution observations. Subsequent studies may consider more realistic accretion disk models, such as geometrically thick disks, and compare black hole shadow images with the optical images of other compact objects, such as boson stars. These efforts may further deepen our understanding of black holes, the most mysterious objects in the universe.

\clearpage
\vspace{10pt}

\noindent {\bf Acknowledgments}
\noindent

This work is supported by the National Natural Science Foundation of China (Grant No. 12375043).

\bibliographystyle{JHEP} 
\bibliography{biblio} 

\end{document}